\documentclass{article}

\usepackage{arxiv}

\usepackage[utf8]{inputenc} 
\usepackage[T1]{fontenc}    
\usepackage{hyperref}       
\usepackage{url}            
\usepackage{booktabs}       
\usepackage{amsfonts}       
\usepackage{nicefrac}       
\usepackage{microtype}      
\usepackage{lipsum}		
\usepackage{graphicx}

\graphicspath{{Image/}{./}} 

\usepackage[inline]{enumitem}
\usepackage{subfig}  
\usepackage{multicol}
\usepackage{multirow}
\usepackage{tabularx}

\usepackage{amsmath,amssymb}
\usepackage{color}
\usepackage{xspace}
\usepackage[table,xcdraw]{xcolor}
\usepackage[normalem]{ulem}
\usepackage{bm}
\usepackage{hyperref}

\makeatletter
\DeclareRobustCommand\onedot{\futurelet\@let@token\@onedot}
\def\@onedot{\ifx\@let@token.\else.\null\fi\xspace}
\def\etal{~et~al\onedot}
\def\eg{e.g\onedot} 
\def\ie{i.e\onedot}

\makeatother

\def\clap#1{\hbox to 0pt{\hss #1\hss}}%
\def\initials#1{\protect\clap{\smash{\raisebox{1.4ex}{\tiny{\textsf{\textit{#1}}}}}}}%
\makeatletter
\newcommand{\EDIT}[4][]{\protect\@ifundefined{hidecomments}{%
  \strut{\color{#3}{\hspace{0pt}\initials{#2}\protect\sout{#1}{~#4}}}%
  }{}}
\newcommand{\NOTEboxed}[3]{\protect\@ifundefined{hidecomments}{%
  {\begin{center}\fbox{\parbox{0.97\linewidth}{\protect\EDIT{#1}{#2}{#3}}}\end{center}}
  }{}}
\makeatletter
\newcommand{\DefAuthor}[2] 
{%
  \expandafter\newcommand\csname #1edit\endcsname[2][]{\protect\EDIT[##1]{#1}{#2}{##2}}
  \expandafter\newcommand\csname #1\endcsname[1]{\protect\csname #1edit\endcsname{[##1]}}
  \expandafter\newcommand\csname #1boxed\endcsname[1]{\NOTEboxed{#1}{#2}{##1}}
}
\makeatother

\newcommand{\abs}[1]{ \left| #1 \right| }

\title{On-Body Visualization of Patient Data for Cooperative Tasks}

\author{
	Dmitri Presnov  \\
	Computer Graphics Group \\
	University of Siegen \\
	Siegen 57068, Germany\\
	\texttt{dmitri.presnov@uni-siegen.de}\\	
	\And
	Julia Kurz 	\\
	Collaborative Research Centre Media of Cooperation \\
	University of Siegen \\
	Siegen 57072, Germany \\
	\texttt{julia.kurz@uni-siegen.de}\\
	\And
	Judith Willkomm \\
	Collaborative Research Centre Media of Cooperation \\
	University of Siegen \\
	Siegen 57072, Germany \\
	\texttt{willkomm@locatingmedia.uni-siegen.de}\\
	\And 
	Daniel Alt \\
	Neurosurgery, Jung-Stilling Hospital \\
	Siegen 57074, Germany \\
	\texttt{Daniel.Alt@diakonie-sw.de} \\
	\And
	Johannes Dillmann \\
	Neurosurgery, Jung-Stilling Hospital \\
	Siegen 57074, Germany \\
	\texttt{Johannes.Dillmann@diakonie-sw.de} \\
	\And
	Robert Zilke \\
	Neurosurgery, Jung-Stilling Hospital \\
	Siegen 57074, Germany \\
	\texttt{Robert.Zilke@diakonie-sw.de} \\
	\And
	Veit Braun \\
	Neurosurgery, Jung-Stilling Hospital \\
	Siegen 57074, Germany \\
	\texttt{Veit.Braun@diakonie-sw.de} \\
	\And
	Cornelius Schubert \\
	Department of Social Sciences \\
	University of Siegen \\
	Siegen 57072, Germany \\
	\texttt{cornelius.schubert@uni-siegen.de} \\
	\And
	Andreas Kolb \\
	Computer Graphics Group \\
	University of Siegen \\
	Siegen 57068, Germany\\
	\texttt{andreas.kolb@uni-siegen.de} \\
}

\begin{document}
\maketitle

\begin{abstract} 
Electronic health records (EHR) systematically represent patient data in digital form. However, text and visualization based EHR systems are poorly integrated in the hospital workflow due to their complex and rather non-intuitive access structure. This is especially disadvantageous in clinical cooperative situations that require an efficient, task specific information transfer.

In this paper we introduce a novel concept of anatomically integrated in-place visualization designed to engage with cooperative tasks on a neurosurgical ward. Based on the findings of our field studies and the derived design goals, we propose an approach that follows a visual tradition in medicine, which is tightly related with anatomy, by using a virtual patient's body as spatial representation of visually encoded abstract medical data. More specifically, we provide a generic set of formal requirements for these kinds of in-place visualizations, we apply these requirements in order to achieve a specific visualization of neurological symptoms related to the differential diagnosis of spinal disc herniation, and we present a prototypical implementation of the visualization concept on a mobile device. Moreover, we discuss various challenges related to visual encoding and visibility of the body model components. Finally, the prototype is evaluated by 10 neurosurgeons, who assess the validity and the further potential of the proposed approach.
\end{abstract}

\section{Introduction}
\label{s:intro}

Hospitals today use electronic health records (EHR) to store patient data in a systematized, digital form and to share it between the different status-groups~\cite{Gunter2005EmergenceNationalelectronic}. In many cases, EHRs are the basis for various automated procedures in hospitals such as clinical decision support, drug procurement and exchange of health information between different stack holders in a hospital~\cite{Menachemi2011BenefitsDrawbacksElectronic}.
Still, EHR systems do not necessarily improve the quality of health care~\cite{Stead2009ComputationalTechnologyEffective}. This is underlined by a recent study, which assessed the physician's beliefs about the meaningful use of the electronic health record~\cite{Emani2017PhysicianBeliefsAbout}. Moreover, EHRs can hardly be used in cooperative situations like shift handovers, as they do not provide sufficient cognitive support~\cite{Staggers2011WhyPatientSummaries}. Such cooperative situations have a very strong focus on \emph{efficient and effective information transfer}. A very recent study documents fundamental problems related to the complex information structure of EHR systems that arise with their long-term usage in cooperative ward rounds~\cite{Wawrzyniak2019ImprovingUsabilitySafety}.

While most EHR systems follow the text-intensive nature of EHRs, there are various approaches that apply methods from information visualization and visual analytics in order to improve clinical support and quality assurance for single patient EHRs or temporal and trend analysis for large EHR databases~\cite{Caban2015VisualAnalyticsHealthcare, West2014InnovativeInformationVisualization}. The majority of approaches in both modes, \ie single patient EHRs and large EHR databases, address the exploration of complete EHR data sets comprising data selection, reconfiguration of layout and visual data encoding, and detection of data correlations or outliers~\cite{Rind2013InteractiveInformationVisualization}.
In clinical practice, however, even simple tasks that require filtering and data collection may be carried out inefficiently by medical practitioners and physicians with standard IT knowledge, due to the complex access structures of information visualization approaches~\cite{Malik2013DoctorsPerspectiveUse}. A recent approach that aims at enhancing the health care practitioners' performance in decision making and care planning by improving the visual data representation in EHR systems confirmed the overall low performance involving EHR systems even if better visual representation are utilized~\cite{Ghassemi2018ClinicalvisSupportingClinical}.

Our long-term interdisciplinary project deals with the fundamental question of analyzing and partly reconfiguring cooperative medical work practices, such as shift handovers and rounds on a neurosurgical ward. Our field studies revealed that
\begin{enumerate*}[label=(\arabic*)]
	\item selected EHR data is indispensable in this kind of cooperative setting,
	\item compared to text-intensive EHR tools, visualization methods are preferable to convey individual patient data in cooperative settings, however
	\item abstract visualization concepts for EHR data are considered as little intuitive and not focused on cooperative tasks.
\end{enumerate*}

There are only very few visualization techniques that potentially allow a more intuitive access to EHR data by being more tightly connected to the visual tradition of clinical practice, \ie by considering the human body as a \emph{canonical spatial representation} for clinical data.
These approaches either use the human body as the central entry point for efficient access and insertion of EHR data items~\cite{Kirby1996PENPADDataEntry}, follow a Google Earth like approach in placing body-referenced information as icons and text in 2D~\cite{Sundvall2007GraphicalOverviewNavigation}, or comprise 3D navigation and visualization methods for EHR data at different levels of detail (LOD) that partially relate to the human anatomy~\cite{An2010LevelDetailNavigation}.
All these approaches are \emph{hybrid} in that they present the final EHR information in textual or abstract form.

In this paper, we represent a novel visualization concept designed to engage with cooperative tasks on a neurosurgical ward, in which an efficient, task specific information transfer is needed.  In contrast to previous hybrid visualization approaches for EHR data, we take advantage of the fact that each specific cooperative task only requires access to a comparatively small subset of EHR data. Therefore, we propose, implement and evaluate a \emph{comprehensive visualization concept} for cooperation relevant neurosurgical medical data. In this regard, this paper contributes to
\begin{enumerate}
	\item The concept of an \emph{anatomically integrated in-place visualization} of abstract medical data including a \emph{formal definition} of requirements for this kind of visualization.
	\item The \emph{design goals} and \emph{data categories} for clinical neurological symptoms related to the differential diagnosis for patients with spinal disc herniation. This results from different field studies and an interdisciplinary iterative design process.
	\item A \emph{prototypical implementation} of the visualization concept and its \emph{evaluation} by neurosurgeons.
\end{enumerate}

\section{Related Work}
\label{s:related}

%
EHR systems based on interactive information visualization are commonly applied in the following application scenarios~\cite{Faisal2012MakingSenseOfHealthInfo}:
\begin{enumerate*}[label=(\arabic*)]
	\item treatment  planning,  
	\item examination  of  patients’  medical records  and their lifelong medical histories, 
	\item representation of pedigrees and family history, 
	\item patient-practitioner communication and shared decision making, and
	\item life management and health monitoring. 
\end{enumerate*}
The patient-practitioner communication relates to designing adequate visual representations when conveying, for instance, risks of cancer. This fundamentally differs from the requirements in cooperative clinical situations addressed by our approach. 
Beyond the application scenario, information visualization based EHR systems are assessed using criteria such as~\cite{Rind2013InteractiveInformationVisualization}:
\begin{enumerate*}[label=(\arabic*)]
	\item data types, \eg options for visualization of categorical and numerical data,
	\item multivariate analysis support, and
	\item number of patient records (one or multiple).
\end{enumerate*}

%
There are several approaches related to the visualization of single patient EHRs that have certain similarity at an abstract level, \eg Plaisant\etal\cite{Plaisant1998LifeLines}, Bui\etal\cite{Bui2007TimeLine}, Craig\cite{Craig2011EHRCollabor}. These approaches commonly use multiple coordinated views (also applied to raw EHR data) to assess an overview of the patient status and medical history, by preventing complex menu and pop-up structures. In these approaches, time-related data are visualized on a timeline.
Craig\cite{Craig2011EHRCollabor} specifically stresses the ``loss of overview'' as one of the main problems of EHR systems and proposes an interface design that resembles classical paper charts. This interface comprises three panes, displaying generic patient information, a timeline with ``events'' related to acquired medical data records, and the specific data related to the currently selected event.
Ghassemi\etal\cite{Ghassemi2018ClinicalvisSupportingClinical} designed an interactive visualization tool for care planning and decision making in intensive care units (ICU) based on existing knowledge about workplace specific practices~\cite{Heath1996DocumentsProfessionalPractice}. They evaluate their tool by comparison to a text-based baseline system. While their approach was slightly superior with respect to accuracy and time-to-decision, the overall task specific performance stayed low.
Belden\etal\cite{Belden2018DesigningMedicationTimeline} designed a medication timeline visualization in the context of ambulatory care of chronic disease. Their iterative user-centered design uses the participatory approach from Sedlmair\etal\cite{Sedlmair2012DesignStudyMethodology} involving workshops with clinical staff; see also Sec.~\ref{s:method}. Related to the specific medication-related task addressed, their evaluation revealed an improved physician performance compared to a text-based presentation.

%
Another thread focuses on the time-related visualization of multiple patients, \eg Guo\etal\cite{Guo2018VisualProgressionAnalysis}, Wang\etal\cite{Wang2009TempSummaries}, Wongsuphasawat\etal\cite{Wongsuphasawat2011LifeFlow}. The main goal of these approaches is related to the analysis of event sequences using methods from visual analytics and statistics in order to find trends and pattern in the patients' treatment history. These analysis methods can also be used to categorize individual patients, \eg using cluster analysis and visualizing  cluster transitions over time~\cite{Guo2018VisualProgressionAnalysis}.
There are only a few approaches that utilize spatial representations in order to provide a more intuitive access to EHR data.
Kirby\etal\cite{Kirby1996PENPADDataEntry} presented one of the first systems that uses a visualization of the human body as the central entry point for efficient access and insertion of EHR data items.
Sundvall\etal\cite{Sundvall2007GraphicalOverviewNavigation} present a prototype of a 2D EHR navigation and visualization framework based on Google Earth and OpenEHR. It supports so-called placemarks, the standard Google Earth approach to position body-referenced information as icons and text.
An\etal\cite{An2010LevelDetailNavigation} developed a 3D navigation and visualization method for EHR data that uses different levels of detail (LOD). The information contained in an EHR is reorganized according to a hierarchical data structure. In this hierarchy, specific user interfaces are assigned to each LOD level that are composed of level-specific graphical entities. Particularly, on the two topmost levels a virtual human body serves as the basis for visualization. Highlighting helps to identify organ systems (level 1) or organs (level 2) affected by some disease. Furthermore, by zooming into the second level, the description of the corresponding disease is represented by text labels and icons.

Based on the insight from our field studies (see Sec.~\ref{s:method.field}) and the design goals deduced from there (see Sec.~\ref{s:method.goals}), we found that none of the existing approaches is suitable for our problem setting. The main gap lies in the lack of efficient and effective information transfer required to tackle specific clinical tasks, which is in agreement with recent studies~\cite{Stead2009ComputationalTechnologyEffective, Emani2017PhysicianBeliefsAbout}.

\section{Design Method and Goals}
\label{s:method}

In the context of our interdisciplinary long-term project on a neurosurgery ward, we investigate fundamental questions of analyzing and in part reconfiguring cooperative medical work practices. 
From a sociological perspective, articulation work~\cite{strauss1985social} is at the core of clinical cooperation, \ie the ongoing work of integrating distributed tasks and maintaining a coherent treatment trajectory by collection, processing and organizing patient data beyond formal divisions of labor.  This information handling is crucial for the physician’s daily routine and mainly relates to the patient's EHR data, which, however, barely supports workflows in cooperative settings.

We aim at the design and development of integrated visual representations of patient information in order to support and potentially even modify specific cooperative workflows.
We address the fundamental problem of the hitherto low acceptance of text- as well as visualization based EHR systems by utilizing a participatory design approach, similar to Belden\etal\cite{Belden2018DesigningMedicationTimeline} and Sedlmair\etal\cite{Sedlmair2012DesignStudyMethodology}. That is, we set up a design and implementation process that involves visualization researchers, sociologists and neurosurgeons in order to analyze specific cooperative real-world situations on the neurosurgical ward.
Our design and implementation process comprises high-frequent interaction between these groups of experts, in particular during the initial definition of the overall goals of the intended visualization tool and the \emph{field studies} that comprise observations of cooperative workflow situations and interviews. The insights of the field study are depicted in Sec.~\ref{s:method.field}, whereas the deduced design goals presented in Sec.~\ref{s:method.goals}.

Moreover, we utilize \emph{participatory refinement of design \& implementation} similar to the principle of agile software development (see, \eg, \cite{Martin2003AgileSW}), in which visualization researchers, sociologists and neurosurgeons jointly advance and refine the visualization design on the basis of the formulated design goals (see Sec.~\ref{s:method.goals}) by utilizing prototype implementations of a visualization tool.
For reasons of efficiency and in order to not bias the evaluation, three physicians (\emph{expert group}) are involved in this stage, whereas the evaluation involves a distinct set of ten physicians (\emph{test group}; see Sec.~\ref{s:eval}).

\subsection{Field Study}
\label{s:method.field}

In this first phase of data collection and processing, field observations and interviews were conducted. Here, the sociologists, and partially visualization researchers, observed neurosurgeons during their routine ward work to become acquainted with the workflows and relevance structures of their daily work. During these observations, specific cooperative work constellations, namely shift handover and ward rounds, were of particular interest.
%
The \emph{main findings} of the field studies are:
\begin{enumerate}[label=\textbf{F\arabic*.}, ref=\textbf{F\arabic*}, itemindent=0pt]
	
	\item \label{enum:findings.coopwork} \emph{Clinical cooperative work} is characterized by the need for efficient and effective information transfer between clinical staff with respect to specific clinical usage contexts in synchronous (\ie face-to-face cooperation, \eg at shift handover) or asynchronous modes (\ie deferred cooperation with absent colleagues, \eg at ward rounds).
	
	\item \label{enum:findings.definciency} A potentially \emph{deficient information transfer} is ascertained in cooperative constellations, as they are dominated by oral, \ie volatile, communication with little information integration from electronic data sources, namely EHR. The main reason is high time pressure and the inefficient access to the relevant EHR data.
	
	\item \label{enum:findings.relevance} There is a strong dependence of the \emph{relevance of an EHR datum} on the clinical usage context. For instance, during a ward round the physician is primarily interested in a patient's neurological status, \eg symptoms related to the damage to sensory nerve fibers.
	
	\item \label{enum:findings.timeline} In addition to the instantaneous state of a patient, the \emph{evolution} of the relevant data is of great importance in order to properly assess the course of the disease or therapy.
	
	\item \label{enum:findings.anormal} In order to allow for an efficient information transfer and to avoid overload and redundancy, physician have a clear \emph{focus on the abnormal} with respect to both, the patient's status as well as the evolution of specific patient data.
	
	\item \label{enum:findings.anatomy} The physicians rely on the \emph{patient's anatomy as spatial reference} not only for diagnosis and surgery planning, using CT images for instance, but also for their articulation work, \eg when presenting a patient in the morning meeting or handwriting information on anatomical sketches in examination forms.
	
	\item \label{enum:findings.vistype} During our semi-structured interviews we asked the clinical staff to assess the potential of various \emph{types of information presentation}. In the process, existing text-based approaches were classified as just as unsuitable as overly abstract visualization concepts.
	
	\item \label{enum:findings.mobil} The physicians need \emph{ubiquitous access} to the clinical data relevant to their cooperative work, \eg on a \emph{mobile device}.
\end{enumerate}

\subsection{Design Goals}
\label{s:method.goals}

Following the project motivation, the overarching goal of the intended visualization system is to facilitate \emph{efficient information transfer in the context of distributed medical tasks}, which has been validated by the field studies (cf. \ref{enum:findings.coopwork}, \ref{enum:findings.definciency}).
The resulting design goals listed below have been jointly deduced by the visualization researchers, sociologists and neurosurgeons from findings of the field studies in Sec.~\ref{s:method.field}:
\begin{enumerate}[label={\textbf{DG\arabic*}},ref={\textbf{DG\arabic*}},itemindent=6pt]
	
	\item \label{enum:dg.familiar} \textbf{Familiarity in novelty.} The goal is to exploit the existing visual tradition, \ie the usage of the anatomy as spatial reference (cf.~\ref{enum:findings.anatomy}) as far as possible and to prevent abstract visualizations (cf.~\ref{enum:findings.vistype}) in order to achieve a high degree of intuitiveness.
	
	\item \label{enum:dg.discriminability} \textbf{Visual discriminability.} Favoring non-abstract visualization concepts increases the problem to properly map data relevant in a given clinical usage context (cf.~\ref{enum:findings.relevance}) in order to attain an effective information transfer (cf.~\ref{enum:findings.definciency}).
	
	\item \label{enum:dg.synopsis} \textbf{Context-related synopsis.} In cooperative settings it is important to provide a simultaneous visual access to all the data that are relevant (cf.~\ref{enum:findings.relevance}) in the given clinical usage context for conveying the current patient status at a glance.
	
	\item \label{enum:dg.intuitive} \textbf{Intuitive comprehensibility.} The visualization design has to focus on an intuitive and comprehensive interpretability in order to address the overall goal of efficient and effective information transfer (cf.~\ref{enum:findings.coopwork} and \ref{enum:findings.definciency}).
	
	\item \label{enum:dg.concise} \textbf{Concise visualization.} In order to avoid clutter and distraction by secondary details, we aim to provide only the data that are relevant in the current clinical usage context (cf.~\ref{enum:findings.relevance}) and to focus on medically relevant, \ie abnormal, value constellations (cf.~\ref{enum:findings.anormal}).
	
	\item \label{enum:dg.progress} \textbf{Visualization of chronological changes.} The visualization should provide access to the evolution of the patient's data (cf.~\ref{enum:findings.timeline}).
	
	\item \label{enum:dg.mobility} \textbf{Mobility.}  The goal is to make the visualization available in all places using a mobile device (cf.~\ref{enum:findings.mobil}).
	
\end{enumerate}

\section{The Visualization Concept}
\label{s:concept}

\subsection{General Considerations}
\label{s:concept.general}

This section describes our visualization concept, which aims at the achievement of the design goals (see Sec.~\ref{s:method.goals}) in consideration of practical insights into the hospital workflow gathered in our field studies (see Sec.~\ref{s:method.field}).

At the core of the concept is an \emph{anatomically integrated in-place visualization} of medical data relevant to specific cooperative tasks. According to this visualization principle, an anatomical model serves as spatial representation of medical data that inherently refer to its structures, \eg the clinical symptom \emph{paresis} referring the affected \emph{muscle}. Particularly, the visual attributes that appropriately encode the abstract data to be visualized are applied by rendering of the corresponding anatomical structures, changing their default or ``natural'' appearance. 

Using the human body as spatial reference, we directly address the design goal \ref{enum:dg.familiar} of exploiting the existing visual tradition, which is tightly bound to the anatomy. On the one hand, this approach reduces the freedom in assigning medical data to visual attributes compared to InfoVis (see, \eg, \cite{Bertin1983SemiologyOfGraphics, Carpendale2003ConsideringVisualVariables}), since the attributes 'spatial position' and 'form' are defined by the model and are no longer free parameters. On the other hand, the focusing on a specific cooperative context and on data relevant therein (cf.~\ref{enum:dg.concise}), significantly reduces the amount of data to be visualized and, consequently, the number of required visual attributes.
This data preselection in combination with in-place visualization provides a context-related synopsis at a glance (cf. \ref{enum:dg.synopsis}). Moreover, we expect that the spatial embedding of abstract medical data in the anatomical context makes its relation to the underlying real phenomena more evident and facilitates its interpretability (cf.~\ref{enum:dg.intuitive}).

However, the intended anatomically integrated in-place visualization poses two main challenges, \ie
\begin{enumerate}
	\item How to \emph{design the mapping} from abstract medical data to visual attributes under the restriction regarding location/geometry, such that visual discriminability (\ref{enum:dg.discriminability}), context-related synopsis (\ref{enum:dg.synopsis}) and intuitive comprehensibility (\ref{enum:dg.synopsis}) can be achieved in practice?
	\item Even though a significant subset of medical data can be inherently related to an anatomical structure, how to deal with medical data that has no canonical relation to a specific anatomical structure, such as a hemogram?
\end{enumerate}
In the rest of the paper, describing our visualization concept, its prototypical implementation and evaluation, we cover medical data \emph{with inherent anatomical reference}. The integration of data without anatomical reference using, for example, more classical methods from InfoVis, will be addressed in future work.

\subsection{Formal Definition and Requirements}
\label{s:concept.formal}

The following provides a formal definition of the visualization problem and the requirements that any specific realization of the anatomically integrated in-place visualization has to fulfill. 

\subsubsection{Medical Data}
\label{s:concept.formal.data}
The medical data to be visualized is structured as follows.
\begin{itemize}
	\item \emph{Property}. A property $p$ is defined as triple
	$$
	p=(\text{propType }, \text{dataType }, \text{domain}),
	$$
	where $p.$propType refers to the underlying medical concept,  $p.$dataType is a nominal, ordinal or numerical data type and $p.$domain is the domain set of elements of the corresponding $p.$dataType. 
	
	\item \emph{Category}. A category $c$ is a pair
	$$
	c=(\text{spatialRef }, \text{props}),
	$$
	where $c$.spatialRef is the reference to the corresponding anatomical structure and $c$.props are all properties of $c$. $C$ is the set of all categories.
	
	\item \emph{View}. A view $V \subseteq C$ is defined as the categories relevant to the given usage context inside the cooperative workflow. $\cal V$ is the set of all views.
	
	Note: Views automatically reduce the amount of information to be visualized, and thus the necessity of user's navigation, by preselecting the context relevant data.
\end{itemize}

The following examples illustrate the structural concept introduced above.

The category `muscle strength' refers to the anatomical structure `muscle', comprising a single property with the type `intensity' that documents the strength a patient can create in a specific muscle, quantified in 6 numerical values.
\begin{align*}
\text{muscleStrength} = &(\text{muscle}, & \\
& \{ (\text{intensity}, \text{numerical}, [0,\ldots,5]) \}.
\end{align*}

The second example describes the category `radicular pain', \ie pain caused by irritation of a nerve root and related to the skin region that is associated with the latter, \ie \emph{dermatome}. This category comprises one numerical and one nominal property.  
\begin{align*}
\text{radicularPain} = (&\text{dermatome}, \\
&\{ (\text{intensity}, \text{numerical}, [0,\ldots,10]), \\
&\phantom{\{}(\text{trigger}, \text{nominal}, \{\text{constant}, \text{stress}) \} \})\mathtt{.}\\     
\end{align*}
There are several fundamental relationships between views, categories and properties:
\begin{itemize}
	\item A property type is unique within a category. Property types obtain their medical meaning only in combination with the category they are used in.
	\item A property type can be shared between properties of several categories. This expresses similarity of medical concepts, such as `intensity' of clinical symptoms in the prior examples, even if the respective property domains can be distinct.
	\item A category is unique within a view.
	\item A category can be shared between several views, as it can be relevant in different usage contexts.
\end{itemize}

In oder to be able to properly specify the mapping of medical data to visual attributes, we define the set $T(c)$ of property types in category $c$ and the set $T$ of all property types as 
\begin{align*}
T(c) &= \{ p.\text{propType}\;|\; p\in c.\text{props}\},\\
T &= \{ p.\text{propType}\;|\; p\in c.\text{props}, c\in C\}.
\end{align*}

\subsubsection{Visual Attributes}
\label{s:concept.formal.visual}
Since shape and position are mainly predefined as the anatomical structures that are used for visualization in our concept, the \emph{geometric} visual attributes for data encoding are restricted to transformations or deformations such that they do not degenerate the respective 3D object. The other available attributes are color components, namely \emph{hue}, \emph{brightness} and \emph{saturation}, \emph{textures} and \emph{transparency}, as well as \emph{time} using, \eg, animations.

Formally, $a$ denotes a visual attribute, $A$ the set of all visual attributes, and $a$.range the discrete and finite set of distinctively perceivable values of $a\in A$. Note, that $\abs{a\text{.range}}$ is commonly smaller than the number of displayable visual attributes.  For example, the visual attribute `hue' is a floating point value, however, the human vision is able to distinguish only up to eight hue values w/o reference (see, \eg, Kuehni\etal\cite{Kuehni2008ColorOrdered}). Similarly, we found that the values of visual attributes `saturation' and `brightness' become hardly distinguishable for a range cardinality of 5 and higher, if no reference color is given in the same scene.  The presence of a visual reference, \eg colorbars, near the visualization can significantly increase the number of distinguishable visual attributes and/or their values. However, displaying of colorbars for all attributes by default would introduce a distraction factor leading to visual clutter. On the other hand, calling these reference colorbars on demand requires as much user interaction as is necessary for accessing raw data by means of textual overlays, whereby the latter provides a more precise information. Thus, we avoid the use of explicit visual references and integrated textual overlays in our approach (see Sec.~\ref{s:concept.further}).

\subsubsection{Visual Encoding / Mapping}
\label{s:concept.formal.mapping}
The mapping of medical data to visual attributes comprises three levels, \ie
\begin{enumerate*}[label=(\alph*)]
	\item encoding of categories,  
	\item encoding of property types, 
	\item encoding of property values. 
\end{enumerate*}
Consequently, the mapping has to allow for the visual retrieval of this information.  The entire mapping process can be represented as:
\begin{enumerate}
	\item Select a visual attribute  $a^C\in A$ to represent categories.
	\item Find a mapping $M^c\colon C \rightarrow a^C.\text{range}$ that maps all categories to values in the range of the selected visual attribute.
	\item Find a mapping $M^t\colon T \rightarrow A\setminus\{a^C\}$ that maps all property types to the remaining visual attributes.
	\item For each $c \in C$ and each $p \in c$.props find a mapping $M^d_p\colon p.\text{domain} \rightarrow (M^t(p.\text{propType}))$.range that maps the property domain values to the range values of the corresponding visual attribute.
\end{enumerate}
We define $M^c$ and $M^t$ on the global set of categories $C$ and property types $T$, respectively, in order to achieve an intuitive comprehensibility according to \ref{enum:dg.intuitive}.

\subsubsection{Injectivity Requirement}
\label{s:concept.inject}
In general, \emph{injectivity} is a pre-requisite for any visual encoding / mapping in order to lead to an unambiguously comprehensible representation (see, \eg, Ziemkiewicz\etal). 
We distinguish the following situations where either the mapping injectivity is strongly required or its violation has to be recognized and appropriately tackled.
\begin{description}[leftmargin=0cm]
	\item[Local Injectivity.] In the case the mapping $M^c$ or $M^t$ is not injective within a given view $V$ or category $c$, respectively, it is impossible to trace back the categories or property types from their visual representation. Local injectivity can be guaranteed if
	\begin{enumerate*}
		\item all categories of a view are mapped to distinct values of $a^C$, and
		\item all property types of a category are mapped to distinct visual attributes.
	\end{enumerate*}
	Formally, this is given, if
	\begin{align*}
	\forall V \in {\cal V}, \forall c \in V &:  M^c(c) \neq M^c(c')\; \forall c' \in V\setminus\{c\},\\
	\forall c \in C, \forall t \in T(c)&:  M^t(t) \neq M^t(t')\; \forall t'\in T(c)\setminus\{t\}.
	\end{align*}
	
	\item[Global Injectivity.] While local injectivity guarantees the visual distinctiveness of categories and properties \emph{inside} each single view and category, respectively, the global injectivity ensures the uniqueness of visual encoding \emph{across} respective contexts, \ie categories and views for property types and views for categories. Formally, we have
	\begin{align*}
	\forall c \in C &:  M^c(c) \neq M^c(c')\; \forall c' \in C\setminus\{c\},\\
	\forall t \in T &:  M^t(t) \neq M^t(t')\; \forall t'\in T\setminus\{t\}.
	\end{align*}
	Due to the restricted distinctiveness of visual attributes, global injectivity is hardly achievable, still it should be pursued as far as possible.
	
	\item[Spatial Injectivity.] Obviously, several categories of a view can refer to the same anatomical structure. Formally, spatial injection in view $V$ is defined as
	$$
	\forall c\in V, \forall c'\in V\setminus\{c\}: c.\text{spatialRef} \neq c'.\text{spatialRef}. 
	$$
	The violation of spatial injectivity rules out simultaneous visualization of the respective medical data and needs to be handled explicitly (see Sec.~\ref{s:concept.further}).
	
	\item[Property Domain Injectivity] A non-injective mapping of a property domain to the respective visual attribute range causes quantization and, consequently, leads to a loss of information. In some cases a quantized visualization can be acceptable, in particular in the context related synopsis (cf.~\ref{enum:dg.synopsis}), in which a qualitative overview is sufficient.  In any case, quantization needs to be detected and reported. Formally, for a given property $p\in c.\text{props},\, c\in C$, quantization can be detected as follows:
	\begin{align*}
	\abs{p.\text{domain}} &> \abs{(M^t(p.\text{propType})).\text{range}}.
	\end{align*}
	In general, quantization cannot be prevented and needs to be handled explicitly (see Sec.~\ref{s:concept.further}).
	
\end{description}

\subsubsection{Visibility Restrictions}
\label{s:concept.visibility}
In general, a 3D visualization with free camera motion intrinsically affects the visibility of geometric objects representing anatomical structures. In our case, we have two causes for restricted visibility.
\begin{enumerate*}
	\item An anatomical structure that relates to relevant medical data (target) may be \emph{occluded} by another anatomical structure. A muscle affected by paresis, for example, is commonly hidden below skin.
	\item The \emph{spatial extension} of an anatomical structure is too small in relation to the entire scene, so that the visualization cannot be clearly recognized. An example would be a tendon.
\end{enumerate*}
Both aspects are handled in our visualization prototype as described in  Section \ref{s:concept.further}.

\subsection{Further Visualization Concepts}
\label{s:concept.further}
Independent from the specific mapping that we introduce in Sec.~\ref{s:implement}, there are further aspects that we added to our anatomically integrated in-place visualization concept in order to achieve the design goals postulated in Sec.~\ref{s:method.goals}.
\begin{description}[leftmargin=0cm]
	\item[Data selection.] Besides two already described data selection mechanisms, which apply automatically, \ie focusing on the abnormal and usage-dependent views, we provide the user with the possibility to additionally \emph{filter the data by their categories}. Note that the usage-dependent preselection reduces the available categories to a manageable amount. This filter allows to tackle, inter alia, the visualization of multiple medical data by means of same anatomical structure, \ie spatial non-injectivity (see Sec.~\ref{s:concept.inject}).
	
	\item[Alternating visualization.] In the constellations where a simultaneous visualization of multiple medical data on the same anatomical structure is not possible, \ie the spatial injectivity is not fulfilled (see Sec.~\ref{s:concept.inject}), an alternating visualization with additional user control to select one of the alternatives, \eg through the data category filter, can be applied (see Fig.~\ref{fig:task1Alt1}-\ref{fig:task1Alt2}). 
	
	\item[Textual overlays.] In some situations, the physician may want to access the underlying information explicitly on demand, \ie in textual form. We enable this by textual overlays on top of the corresponding anatomical structure in order to, for example, resolve the quantization problem (see Sec.~\ref{s:concept.inject}) that results in a simplified visual representation, or to support the physicians in the getting acquainted with our visualization tool.
	
	\item[Proxies.] The small object extension problem (see Sec.~\ref{s:concept.visibility}) can be tackled by means of an appropriately scaled proxy that is projected to the body surface over the location of the target anatomical structure and rendered with the corresponding visual attributes.
	
	\item[Transparencies.] In order to handle depth occlusions of anatomical structures carrying relevant information by other anatomical structures (see Sec.~\ref{s:concept.visibility}), we use view dependent (semi\nobreakdash-)transparency for the occluder, in case the occluder itself is not carrying relevant information, while preserving the surrounding anatomical context.
	
\end{description}

The specific technical approaches taken to implement textual overlays, proxies and transparencies are described in Sec.~\ref{s:implement.features.details}.

\newcommand{\qunt}{$^1$}
\newcommand{\prxy}{$^2$}
\newcommand{\cmp} {$^3$}

\begin{table*}[tb!]
	\centering
	\renewcommand{\arraystretch}{1.1}
	\caption{Data Categories: The raw data categories (left block), the final categories after discussion with the physicians (center block), and the visual attributes incl. anatomical reference (right block). Only the abnormal states are listed. Specific aspects are indicated as \qunt: quantization, \prxy: usage of proxy geometry, \cmp: extended range due to explicit comparison.}
	\label{tab:dataCat}
	\begin{tabular}{|p{.08\textwidth}|p{.08\textwidth}|p{.08\textwidth}||p{.08\textwidth}|p{.08\textwidth}|p{.08\textwidth}||p{.16\textwidth}|p{.14\textwidth}|} \hline
		\multicolumn{3}{|c||}{\bfseries Raw Data Categories} & \multicolumn{3}{c||}{\bfseries Final Data Categories} & \centering\textbf{Visual Attribute} & \textbf{Anatom. Reference} \\\cline{1-6}
		\centering\textbf{Category} & \centering\textbf{Prop. Type} & \centering\textbf{Domain}& \centering\textbf{Category} & \centering\textbf{Prop. Type} & \centering\textbf{Domain}& \centering\textbf{( \rule[-0.3mm]{0.3mm}{2.4mm} range \rule[-0.3mm]{0.3mm}{2.4mm}~)}& \\
		\hline\hline
		\multicolumn{3}{|l||}{Radicular Pain}     & \multicolumn{3}{l||}{Radicular Pain}     & Red                      & \\\cline{2-3}\cline{5-6}
		& Intensity  & $\{1, \ldots, 10\}$        & & Intensity   & $\{1, \ldots, 10\}$      & Saturat.-Brightn.~(3)\qunt  & Dermatome \\\cline{2-3}
		& Trigger     & binary                     & & Trigger      & binary                   & Texture~\emph{Normal Pert.}(1)     & \\\hline
		\multicolumn{3}{|l||}{Muscle Strength}    & \multicolumn{3}{l||}{Paresis}            & Purple                   & \\\cline{2-3}\cline{5-6}
		& Intensity  & $\{1, \ldots, 5\}$         & & Intensity   & \{mild, moderate, severe\}& Saturat.-Brightn.~(3)   & Muscle \\\hline
		\multicolumn{3}{|l||}{T-Reflex}           & \multicolumn{3}{l||}{T-Reflex}           & Green                    & \\\cline{2-3}\cline{5-6}
		& Intensity  & $\{1, \ldots, 5\}$         & & Intensity   & $\{1, \ldots, 5\}$       & Saturat.-Brightn.~(5)\cmp & Tendon\prxy \\\hline
		\multicolumn{3}{|l||}{Excretion Disorder} & \multicolumn{3}{l||}{Excretion Disorder} & Orange                   & \\\cline{2-3}\cline{5-6}
		& Intensity  & binary                     & & Intensity   & binary                   & Saturat.-Brightn.~(3)    & Urethra or anus\prxy \\\hline
		\multicolumn{3}{|l||}{Paresthesia}        & \multicolumn{3}{l||}{Sensory Disorder}   & Cyan                     & \\\cline{2-3}\cline{5-6}
		& Intensity  & $\{1, \ldots, 3\}$         & & Intensity   & $\{1, \ldots, 4\}$       & Saturat.-Brightn.~(4)    & Dermatome \\\cline{1-3}
		\multicolumn{3}{|l||}{Hypoesthesia}       & & Paresthesia & $\{1, \ldots, 3\}$       & Texture~\emph{Noise}~(3) & \\\cline{2-3}\cline{5-8}
		& Intensity  & $\{1, \ldots, 3\}$         & \\\cline{1-3}
		\multicolumn{3}{|l||}{Anaesthesia}        & \\\cline{2-3}
		& Intensity  & binary                     & \\\cline{1-4}
	\end{tabular}
\end{table*}

\section{Prototype Implementation}
\label{s:implement}

%
Based on the design goals stated in Sec.~\ref{s:method.goals} and the visualization concept introduced in Sec.~\ref{s:concept}, we present a prototype implementation on a mobile device.
As a proof of concept, we focus on patients with \emph{spinal disc herniation}. Moreover, we concentrate on clinical symptoms that are relevant for the \emph{retrieval of a patient's neurological status}, which turns out to be the common usage context for the asynchronous cooperative situations such as ward rounds (cf. \ref{enum:findings.coopwork}) and defines the main view of our prototype.

\subsection{Development Environment}
\label{s:prototype.env}
As geometric model for the anatomically integrated visualization we use \emph{plasticboy's} anatomical 3D human avatar\footnote{\protect\url{www.plasticboy.co.uk/store/Human_Male_Female_Anatomy_Complete_V05.html}}. The model already includes the main organ systems subdivided into the corresponding anatomical structures. However, the granularity of the hierarchical anatomic structure is partially too coarse and has to be extended in order to allow for a proper spatial mapping. Particularly, in the context of neurosurgery, the dermatomes constitute very important anatomical structures that are not reflected in the model.
Hence, we have set up an anatomy refinement procedure based on indexed texture maps that is applied to the existing 3D geometries, \ie polygonal meshes. Our procedure comprises 3D painting functionalities commonly available in modeling tools such as Maya. By ``painting'' the required anatomical sub-structure on the geometry of the (super-)structure and saving the resulting segmentation as indexed texture, these index textures can be used for looking up the anatomical sub-structures of the current fragment in the fragment shader.

Our rendering framework uses the Vulkan API under Android allowing an efficient resource management, which is important especially on mobile devices. In order to fit all high resolution model textures to the limited mobile graphics memory, we initially convert them into the ASTC format~\cite{Nystad2012AdaptiveScalableTexture}.

\subsection{Prototype Features}
\label{s:implement.features}

\subsubsection{Mapping of Spinal Disc Herniation Data}
\label{s:implement.features.mapping}
\begin{description}[leftmargin=0cm]
	\item[Raw Data Categories.] In collaboration with the neurosurgeons, we defined seven data categories (see \autoref{tab:dataCat}) that are of high relevance with respect to the representation of neurological status of a patient with spinal disc herniation, \ie in the respective view (cf. Sec.~\ref{s:concept.formal.data}). The type of the main property to be visualized, common for all these categories, is the \emph{intensity} with which the respective symptom manifests, whereas its domain is individual for each symptom, \ie category.

	The category \emph{radicular pain}  has a further property with the type \emph{trigger}, which states if the pain is constant or only occurs under stress, \eg during movements, whereby the former is assumed to be the normal, \ie default situation for pain that does not need any visual indication.
	
	\item[Data Category Refinement.] During the first trials with the prototype and discussions with neurosurgeons, two changes to the initial raw data categories have been applied.
	First, the three categories related to \emph{sensory disorder}, \ie paresthesia, hypoesthesia and anaesthesia have rather complex interrelations. For example, hypoesthesia and anaesthesia can be considered as different stages of sense decrease, whereas paresthesia is in a certain sense orthogonal because it does not describe a decrease of sensation but rather its abnormality, \eg tingling, and, thus, it can occur in combination with hypoesthesia. Therefore, the new category sensory disorder represents anaesthesia and hypoesthesia as a joint property with the type `intensity' and has an additional property with the type `paresthesia'. 
	Second, insufficient \emph{muscle strength}, measured in the Medical Research Council scale $0,\ldots,5$, is used in daily clinical practice as indication for \emph{paresis} that trigger for potential urgent actions such as emergency surgery. Therefore, we adopted this practice by using the category \emph{paresis} with the ordinal data type comprising the values `mild', `moderate', `severe'.

	\item[Mapping of Data to Visual Attributes.] The final medical data categories in \autoref{tab:dataCat} are visually encoded in consideration of the rules deduced in Secs.~\ref{s:concept.formal.mapping} and \ref{s:concept.inject}, that is, the mapping functions $M^c$ and $M^t$ are at least locally injective. We selected \emph{hue} as the visual attribute $a^C$ to encode \emph{category}. The category mapping $M^c$ takes into account the distinctiveness of the resulting five hues with regard to each other as well as to their context in the anatomical model, \eg a red color is not suitable for visualization on muscles because it highly coincides with their natural, \ie healthy, appearance. The shared property type \emph{intensity} is mapped in all categories to a composite visual attribute \emph{saturation-brightness} in the HSV color space, \ie $M^t$ is also globally injective. We decided to combine two respective visual attributes in a single one to increase the visual discriminability (\ref{enum:dg.discriminability}) especially in cases of the visualization of a single property value that is difficult to assess if no color reference, \eg colorbar, is present (see discussion in Sec.~\ref{s:concept.formal.visual}). \emph{Additional properties} are encoded by means of \emph{textures}. We select texture such that they are visually as complementary to the visual attribute color as possible in order to allow for a simultaneous visualization. In case the additional property is non-binary, \eg the property with the paresthesia type in sensory disorder (cf. Sec.~\ref{tab:dataCat}), the required range of visual values corresponds to the texture's frequency and amplitude; see also~Fig.~\ref{fig:legend}.
	
	There are several features of the mapping to visual attributes that need to be mentioned (cf.~\autoref{tab:dataCat}):
	\begin{enumerate*}[label=(\arabic*)]
		\item The 10 intensities of the pain category result in a visual quantization, \ie the corresponding property domain mapping is not injective (cf. Sec.~\ref{s:concept.inject}),
		\item the anatomical reference for the categories \emph{T-reflex} and \emph{excretion disorder} are too small and require a proxy geometry (see also Sec.~\ref{s:implement.features.details} and, \eg, Fig.~\ref{fig:timeline}), and
		\item we can slightly increase the saturation-brightness cardinality by $1$ for the main property of the category \emph{T-reflex} as abnormal intensity values occur in anatomical pairwise constellations, \eg on the left and right leg and the main indication is their difference.
	\end{enumerate*}

	
	

\end{description}

\subsubsection{Rendering Implementation Details}
\label{s:implement.features.details}
\begin{figure}
	\centering
	\subfloat[]{
		\includegraphics[width=0.24\textwidth]{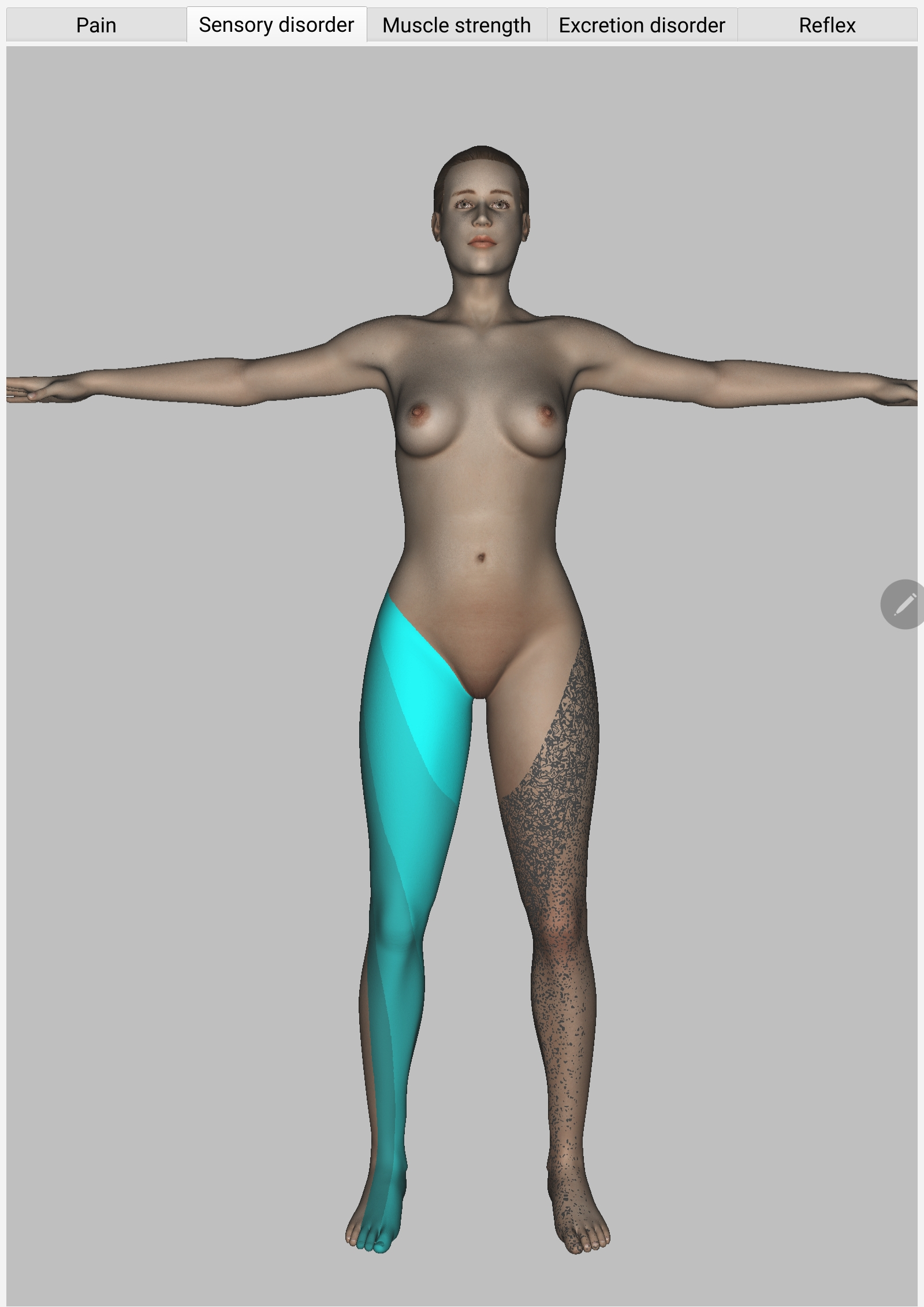} 
		\label{fig:legend_sensory}		
	} 
	\subfloat[]{
		\includegraphics[width=0.24\textwidth]{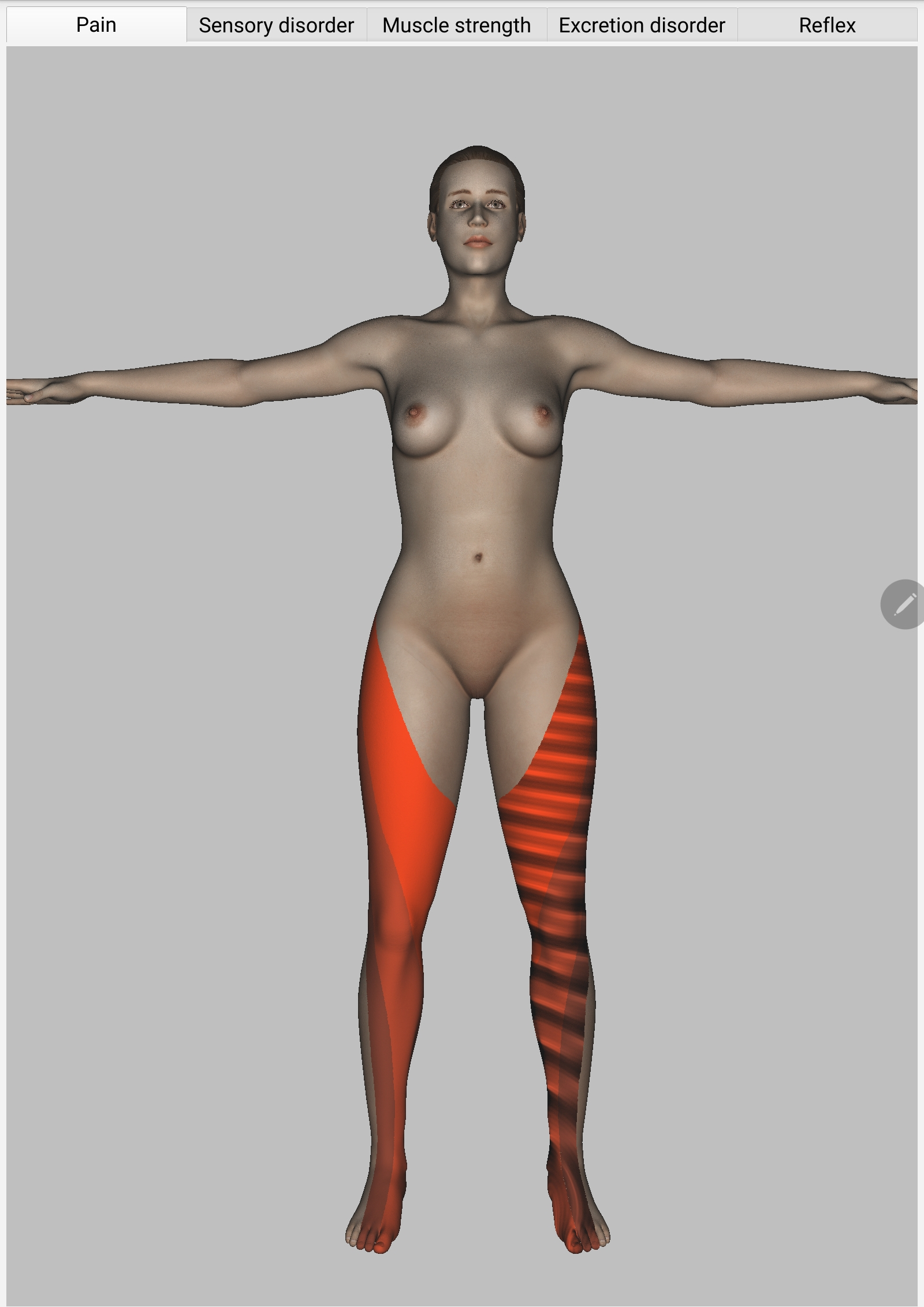}
		\label{fig:legend_pain}		
	} 
	\caption{The legend of the prototype visual encoding. Fig.~\ref{fig:legend_sensory} depicts the active tab of the sensory disorder category: the right L2-L5 dermatomes visualize four intensity levels, the left L3-L5 dermatomes show three levels of the paresthesia, \ie Noise texture; Fig.~\ref{fig:legend_pain} shows the active radicular pain tab: the right L3-L5 dermatomes visualize three intensity levels, the left L3-L5 dermatomes show same levels in combination with the stress trigger, \ie Normal Pert. texture (cf. Tab.~\ref{tab:dataCat}); the exact property values can be looked up by means of overlays similarly to Fig.\ref{fig:task1Overlay}.}
	\label{fig:legend}	
\end{figure}

\begin{figure*}
	\center
	\subfloat[]{
		\includegraphics[width=0.32\textwidth]{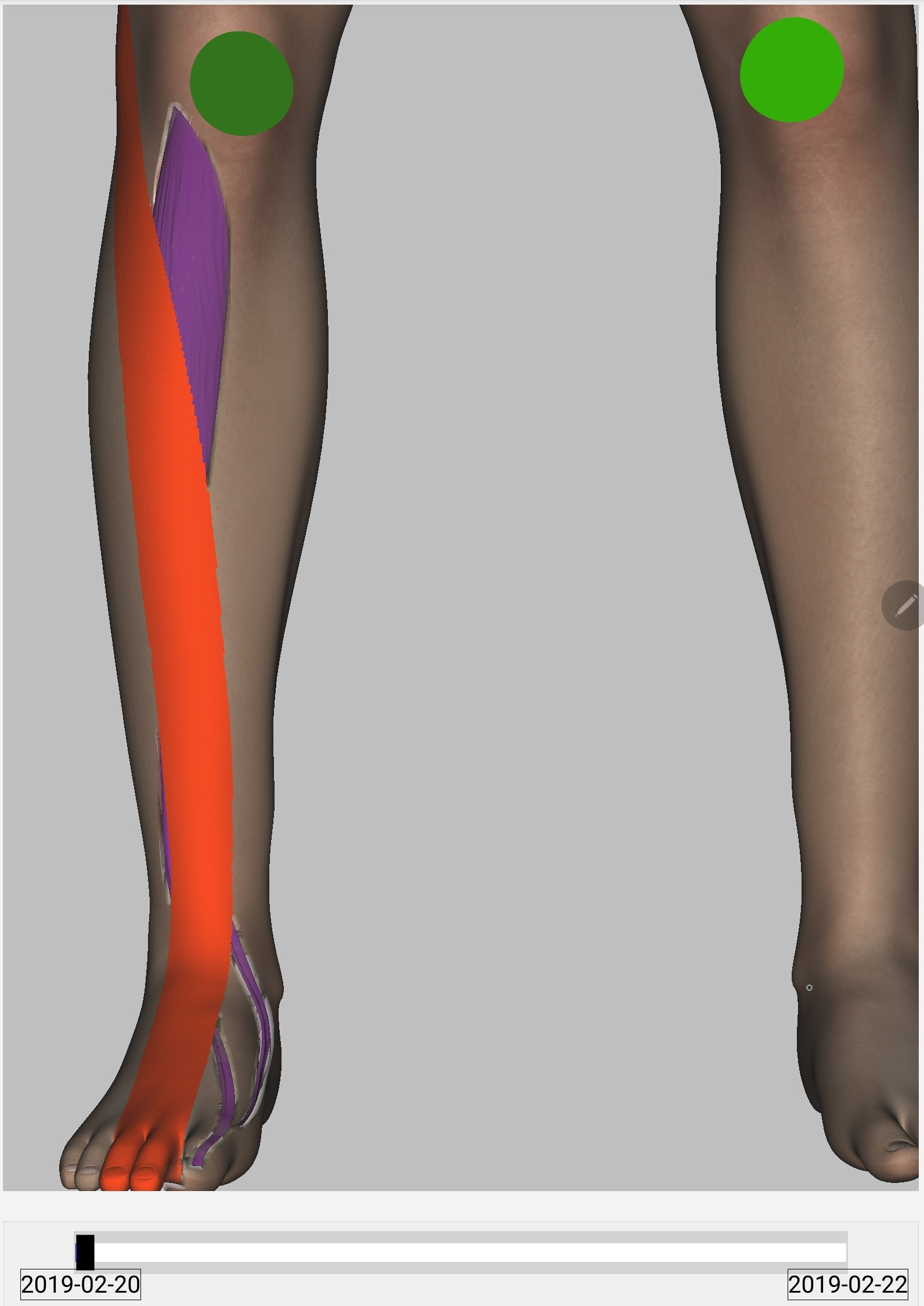} 
		\label{fig:task1Alt1}
	} 
	\subfloat[]{
		\includegraphics[width=0.32\textwidth]{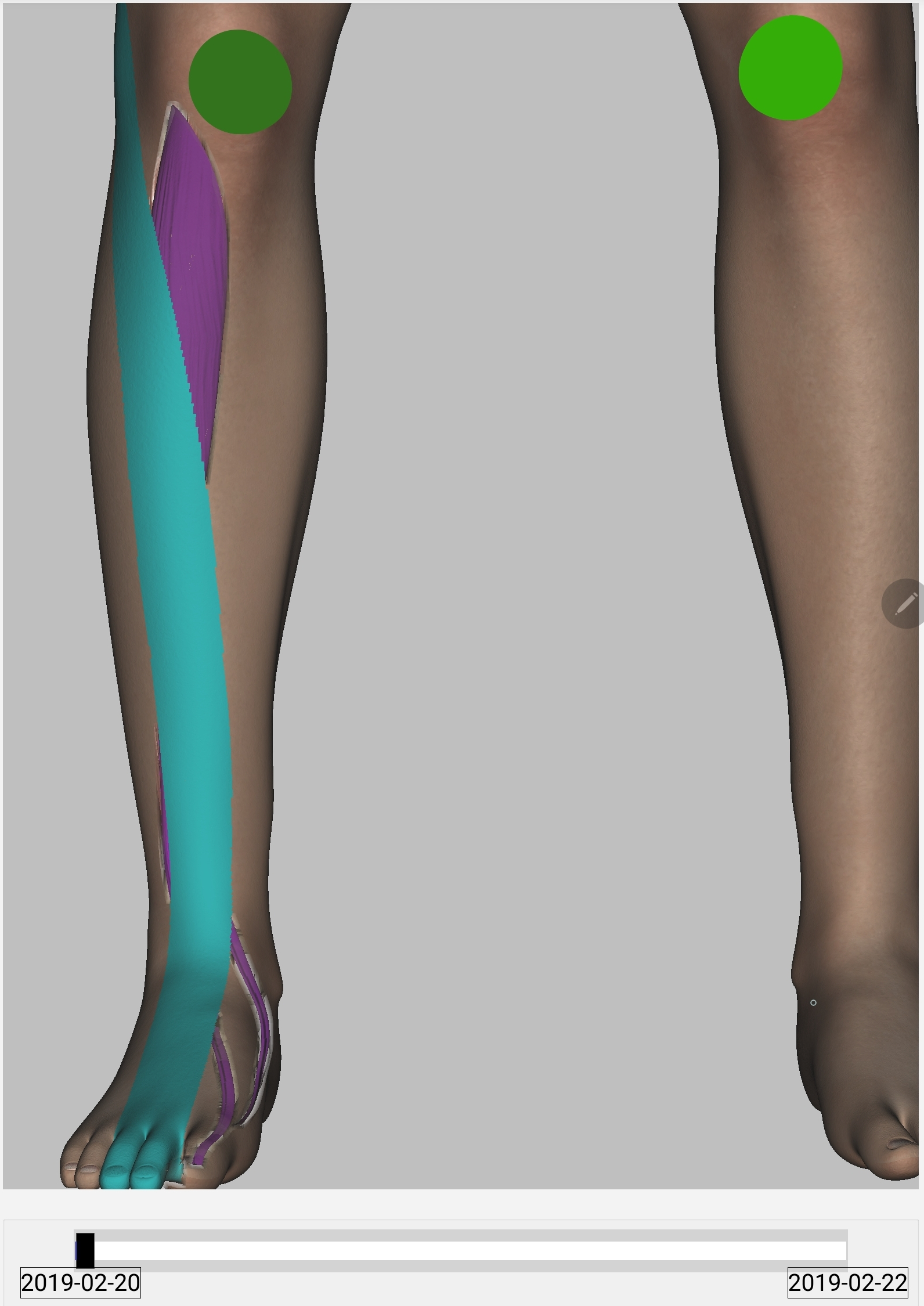} 
		\label{fig:task1Alt2}
	} 
	\subfloat[]{
		\includegraphics[width=0.32\textwidth]{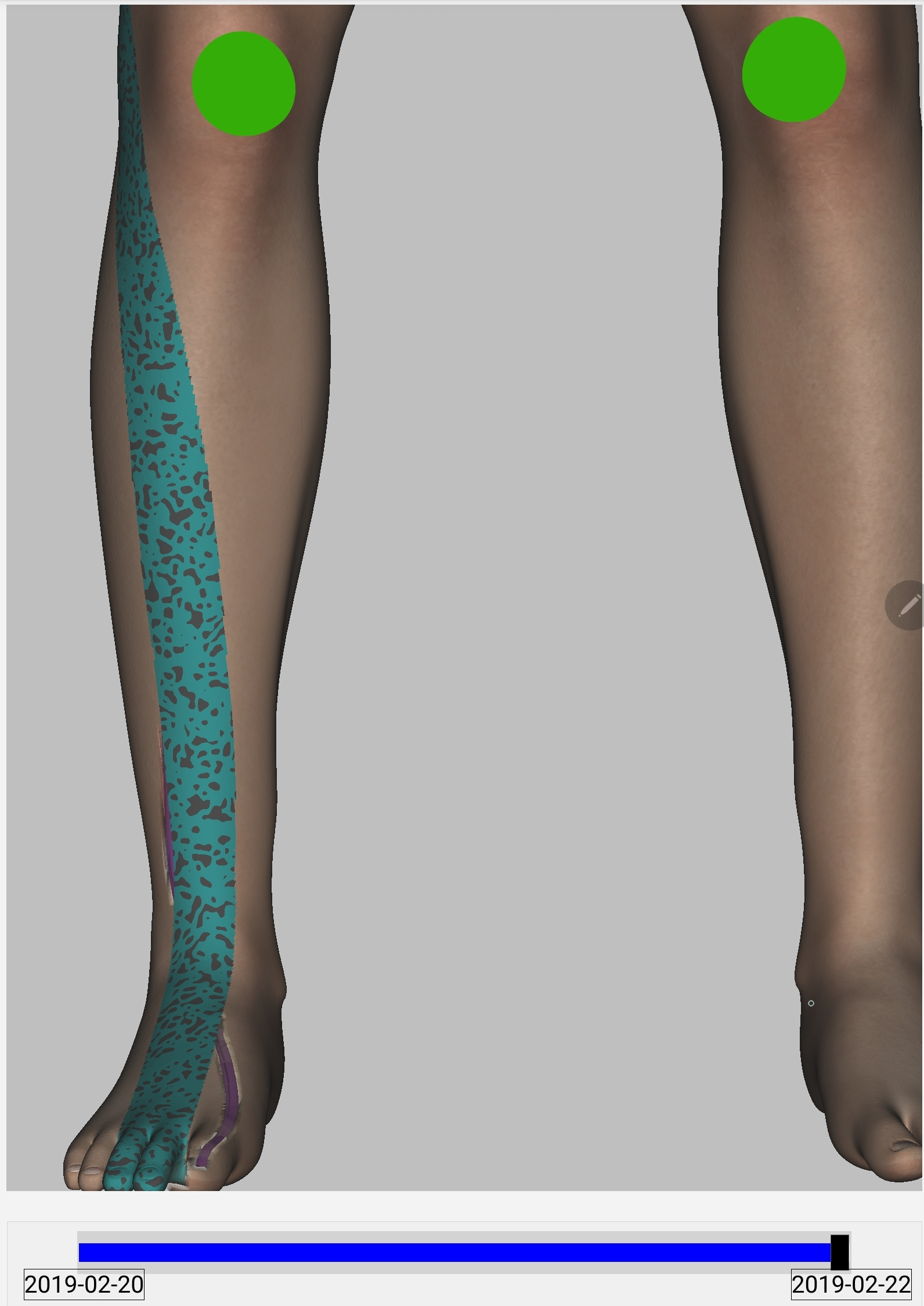} 
		\label{fig:task2Sensory}
	}
	\caption{Visualization of temporal changes. Fig.~\ref{fig:task1Alt1}-\ref{fig:task1Alt2} show the visualization of the preoperative neurological status of a patient with an asymmetric patellar reflex, a moderate paresis in the right extensor hallucis longus muscle, a mild paresis in the tibialis anterior muscle, a severe pain and a hypoesthesia in the left L5 dermatome, whereby the pain (see Fig.~\ref{fig:task1Alt1}) and the hypoesthesia (see Fig.~\ref{fig:task1Alt2}) are visualized alternately; this visualization is used in \ref{enum:task2}. Fig.~\ref{fig:task2Sensory} shows the postoperative status of the same patient with symmetry in patellar reflex, a decreased hypoesthesia in combination with paresthesia and a remaining mild paresis in the right extensor hallucis longus muscle (see \ref{enum:task3}).}
	\label{fig:timeline}
\end{figure*}

In this section, we briefly describe some implementation details related to the realization of specific visualization features, partially already discussed in Sec.~\ref{s:concept.further}.
\begin{description}[leftmargin=0cm]
	\item[3D Solid Textures.] The interference of the textures, applied for visualizing additional data properties, with the color components hue (visual attribute to encode categories, $a^C$) and saturation-brightness (property type \emph{intensity}) that are already assigned to the same anatomical structure should be as low as possible (cf. \ref{enum:dg.discriminability}). This is achieved by using textures that either sparsely vary the intensity or create lighting effects by means of a normal perturbation, \ie procedural normal maps.
	The texture coordinates provided by the 3D human body model refer to texture atlases, which introduce geometrical distortion and whose segmentation does not follow the anatomical structures, \ie we have no means of applying textures directly to the given coordinates. Therefore, we use 3D solid textures and the vertex positions as texture coordinates to achieve a proper and distortion-free texture parametrization~\cite{Pietroni2010SolidTextureSynthesis}. In particular, we chose the procedural form of solid textures that allows to synthesize fragment color on the fly in a shader without any consumption of GPU global memory on our mobile device, which has limited hardware resources. The examples in \autoref{tab:dataCat} (see Fig.~\ref{fig:legend} and Fig.~\ref{fig:task2Sensory}) are stochastic textures based on Perlin's noise (Texture~\emph{Noise}) and on normal perturbation using cycloidal functions (Texture \emph{Normal Pert.}) (see ~\cite{Perlin1985ImageSynthesizer,Hart1999AntialiasedParameterizedSolid}).
	
	\item[Anatomical Proxies.]
	The main idea in generating anatomical proxies is to utilize projective textures~\cite{Segal1992FastShadowsLighting}, \eg appropriately scaled circles, on the skin surface above the anatomical structure that is too small for a direct visualization (cf. Sec~\ref{s:concept.visibility} and the patellar reflex, \eg, in Fig.~\ref{fig:timeline}).
	To achieve this, several pre-processing operations are performed. Firstly, we compute a principle component analysis (PCA) of the target anatomical structure on the GPU using the vertex data stored in buffer objects in graphics memory, thus preventing unnecessary data transfer. The PCA's center point serves as lookat point of the projection, whereas the PCA's main axes are used to realize optional shifts or scales of the proxy consistently with the object's extent, \eg in the case of the triceps tendon reflex the lookat point is shifted to the muscle attachment on the elbow.
	Secondly, we calculate the minimum and maximum skin depth values in projector space by rasterizing the avatar's skin into two depth buffers using a fragment shader and a subsequent minimum and maximum finding in a compute shader. In order to restrict the texture projection to the avatar's skin surface closest to the anatomical structure and to prevent projection onto the avatar's far side, we use the midpoint of the minimum and maximum depth value inside the projector footprint as depth threshold in projection space and discard all fragments that are greater or lesser, depending on whether the target anatomical structure lies near to the front or back body side, respectively.

	\item[Occlusion Handling]
	In order to visualize hidden anatomical structures in an integrated overview of the most relevant information without requiring specific navigation or zooming efforts, we dynamically decrease the opacity of areas above the occluded target structure, depending on the current camera transformation (cf. Sec.~\ref{s:concept.visibility}). Similar to Viola\etal\cite{Viola2004ImportanceDrivenVolume} and Burns\etal\cite{Burns2008AdaptCut}, we use an image space approach that allows an efficient detection of occluding fragments in real time on mobile hardware. First, all anatomical structures that carry information are rendered in an offline step creating their footprint in the depth buffer, which is implemented as Vulkan storage image. Subsequently, the opacity of the fragments that are nearer to the camera than the corresponding depth value in the footprint is modified. In our approach we set the opacity close to zero in order to prevent interference with the color of the underlying structures that might degrade a proper recognition of referenced medical data. Furthermore, we extend the transparency region by a margin with increasing opacity depending on the pixel distance to the footprint edge; this improves the visible perception of the transparent cutouts.
	
	For a better depth perception, the anatomical hierarchy is also taken into account during occluder detection. In particular, only anatomical structures that are part of the same body region are considered as potential occluders, \eg the left tibialis anterior muscle can be seen through the structures of the left lower leg but not through the right foot (see Fig.~\ref{fig:occlusion}). For this purpose, we additionally store the body region's ID in the appropriately extended depth buffer and compare it at rendering time in the fragment shader with the ID of the potential occluder.
	
	Finally, the (semi-)transparent fragments are blended into the framebuffer image. Here, the correct rendering order is crucial. Therefore, we sort the fragments in image space in front-to-back order using a depth peeling approach~\cite{Everitt2001OrderTransp}. In order to reduce the number of render passes, we use the Vulkan subpass concept combining respective peeling and blending steps.
	
	We accelerate the above algorithm by extending the depth/region buffer with a bit mask generated in the first depth peeling pass that masks out image regions that do not contain target objects, \ie anatomical structures that carry medical information.  The pixels that do not overlay the footprint with its extended margin will not become transparent fragments and will be discarded in all subsequent peeling and blending passes.

	\begin{figure}[h]
		\centering
		\includegraphics[width=0.4\textwidth]{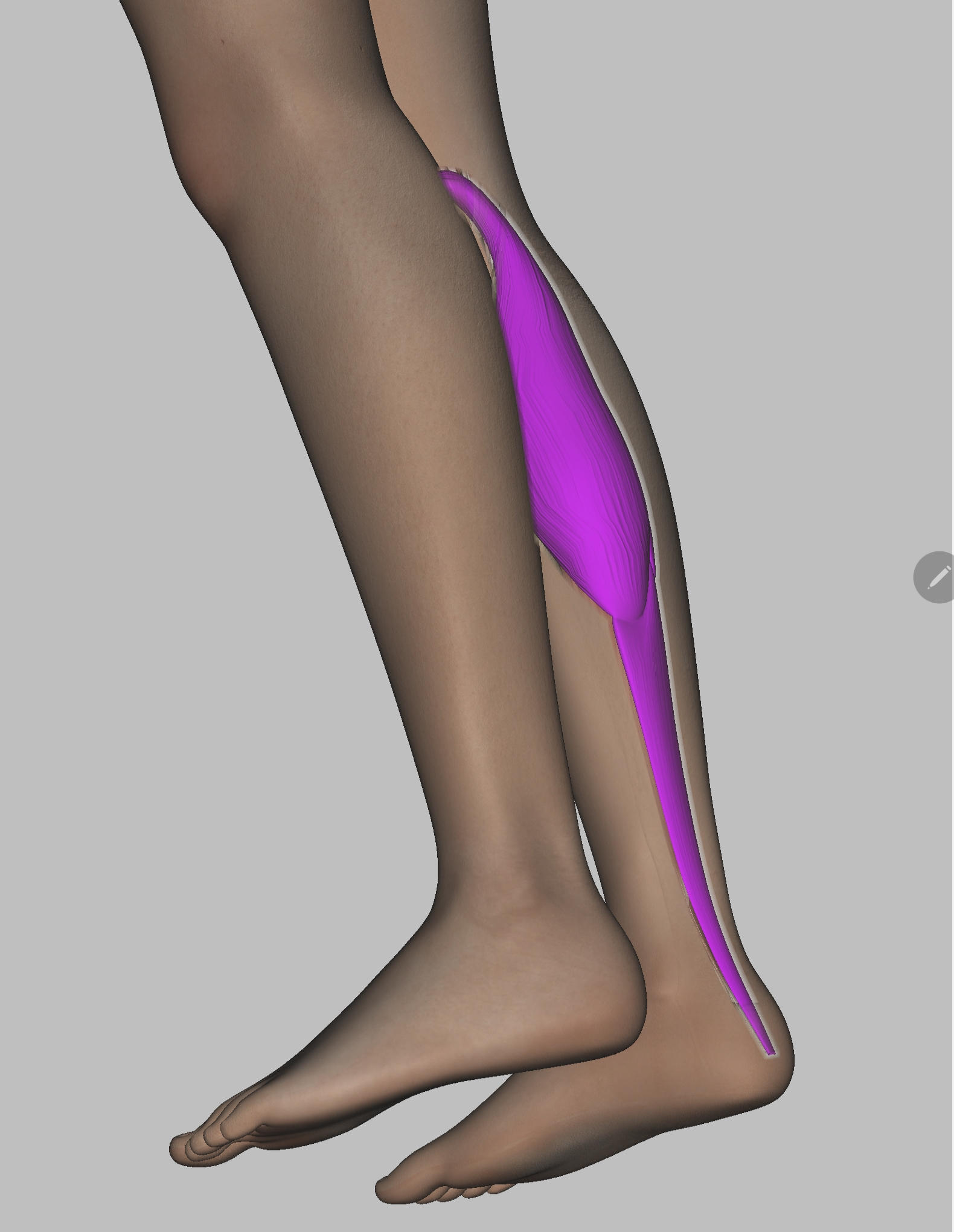}
		\caption{Dynamic transparency area with hierarchical information: the right gastrocnemius muscle with paresis data (purple) is visible through the skin of the same body region and partially occluded by the left leg.}
		\label{fig:occlusion}
	\end{figure}
	
	\item[Picking and Context Menu] The prototype includes a context menu for accessing the advanced features such as the \emph{data category filter} (see Fig.~\ref{fig:task1Filter}) and the \emph{overlays} with textual data (see Fig.~\ref{fig:task1Overlay}). The filter provides for the user the possibility to hide/show data visualizations by their category as described in Sec.~\ref{s:concept.further}, \eg for handling of the spatially non-injective cases (see Sec.~\ref{s:concept.inject}). The overlays display textual data corresponding to the visualization on a anatomical structure selected by user. This, in turn, requires an appropriate mechanism for picking of 3D objects. In our prototype the picking is realized in the image space and is combined with the offline step of the occlusion handling described above. More precisely, the ID's of target anatomical structures are written in the extended depth buffer during occlusion handling. The ID of the current pixel position can be read out, copied back to the CPU memory and used for retrieval of the associated data.
	
	\item[Visualization of Temporal Changes]
	For assessment of healing progress a slider with dates of available clinical examinations is integrated below the main 3D view (see Fig.~\ref{fig:timeline}). Moving the slider, the user can navigate to the date of interest or scroll consecutive examination results. By selecting a date, the entries with the corresponding timestamp are retrieved from the database and the virtual body is rendered with updated visual attributes.
	
\end{description}

\begin{figure}[h!]
	\centering
	\subfloat[]{
		\includegraphics[height=0.24\textheight]{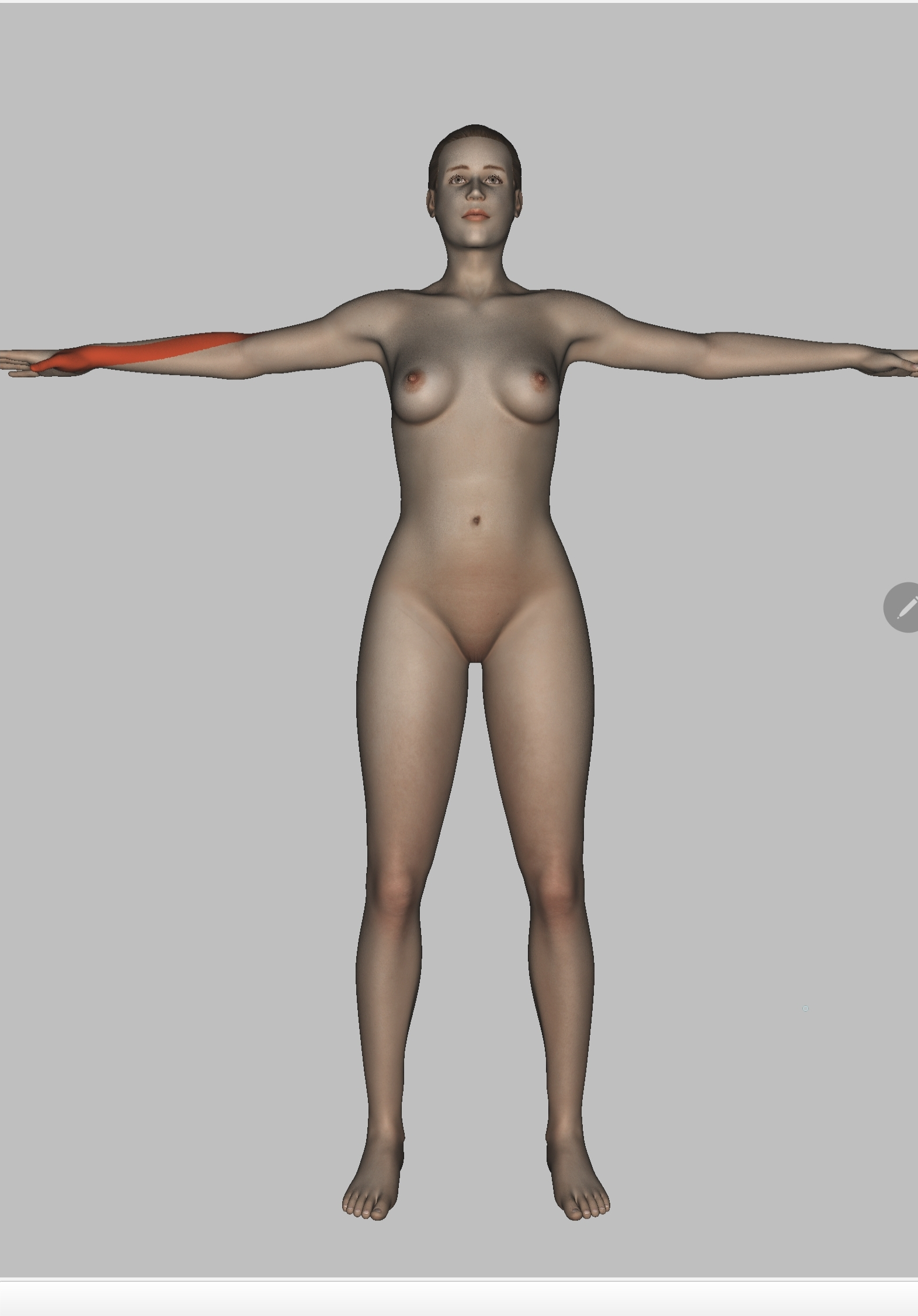} 
		\label{fig:task0_1}
		
	} 
	\subfloat[]{
		\includegraphics[height=0.24\textheight]{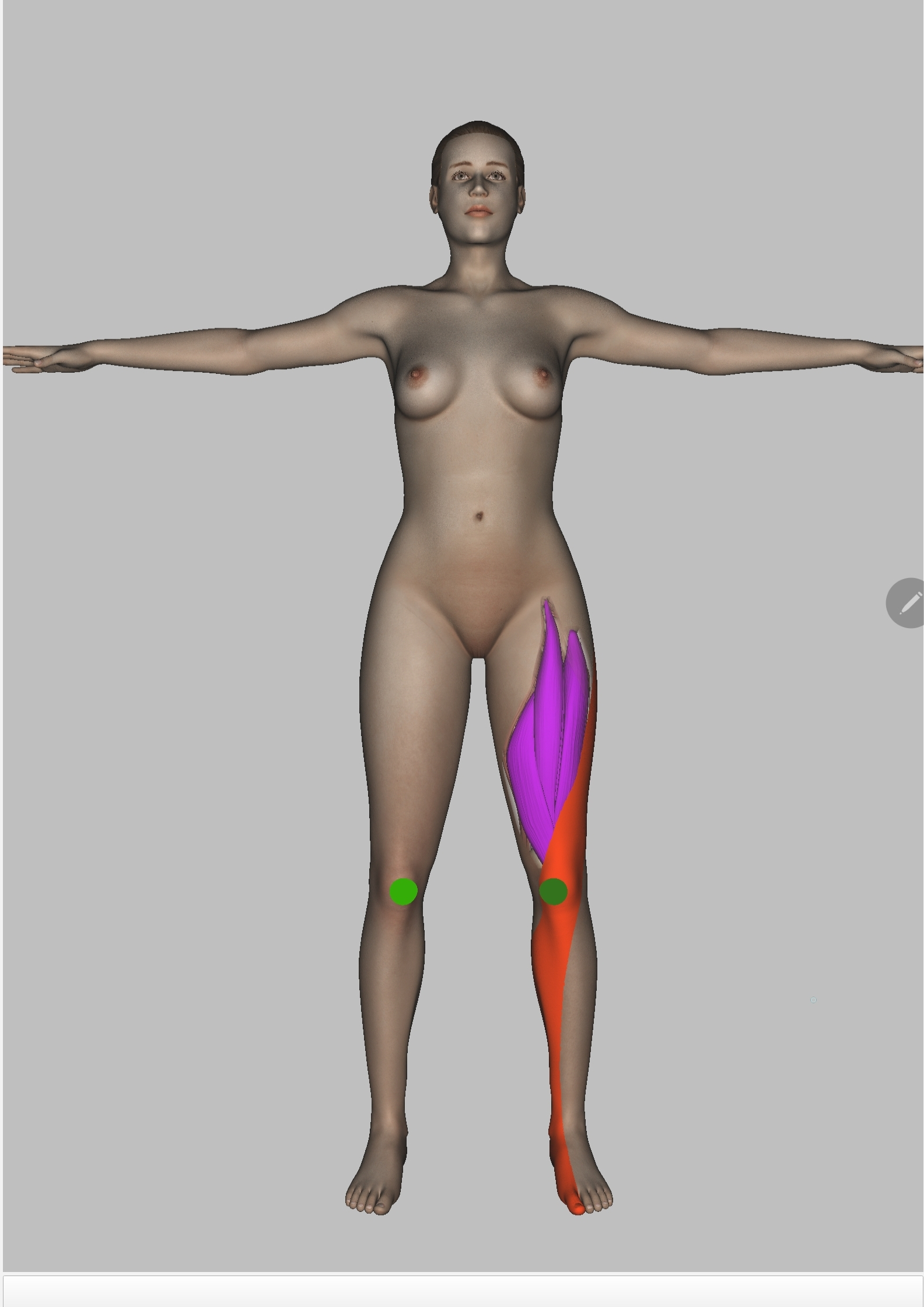} 
		\label{fig:task0_2}
		
	} 
	\caption{Two visualization samples used for spontaneous interpretation in \ref{enum:task1}. Fig.~\ref{fig:task0_1} shows right C6 dermatome with radicular pain. Fig.\ref{fig:task0_2} shows several symptoms visualized in parallel: radicular pain in the left L4 dermatome, paresis of the left quadriceps muscle, asymmetric patellar reflex by means of the proxies.}
	\label{fig:task0}
\end{figure}

\begin{figure*}
	\center
	\subfloat[]{
		\includegraphics[width=0.32\textwidth]{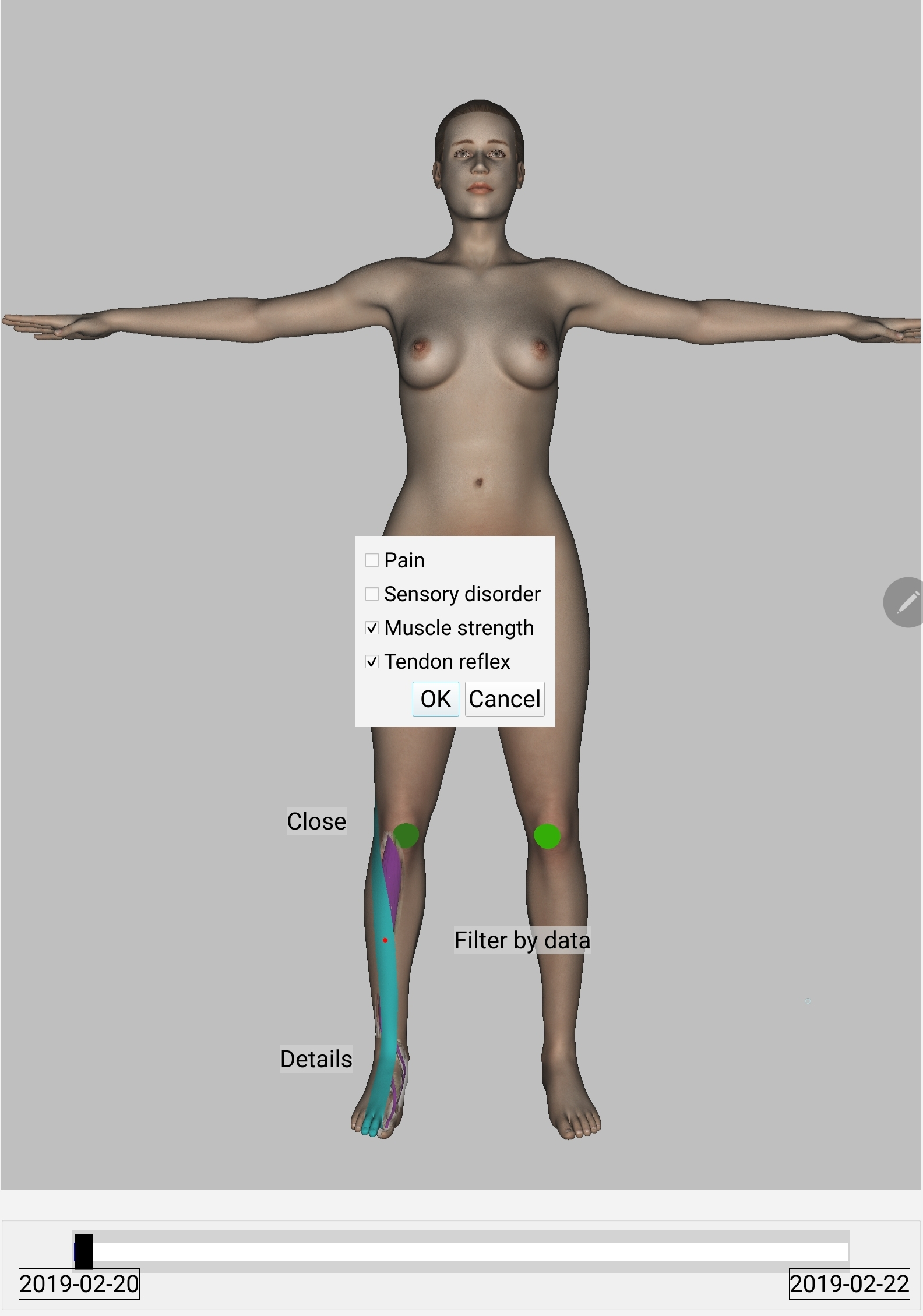} 
		\label{fig:task1Filter}
	} 
	\subfloat[]{
		\includegraphics[width=0.32\textwidth]{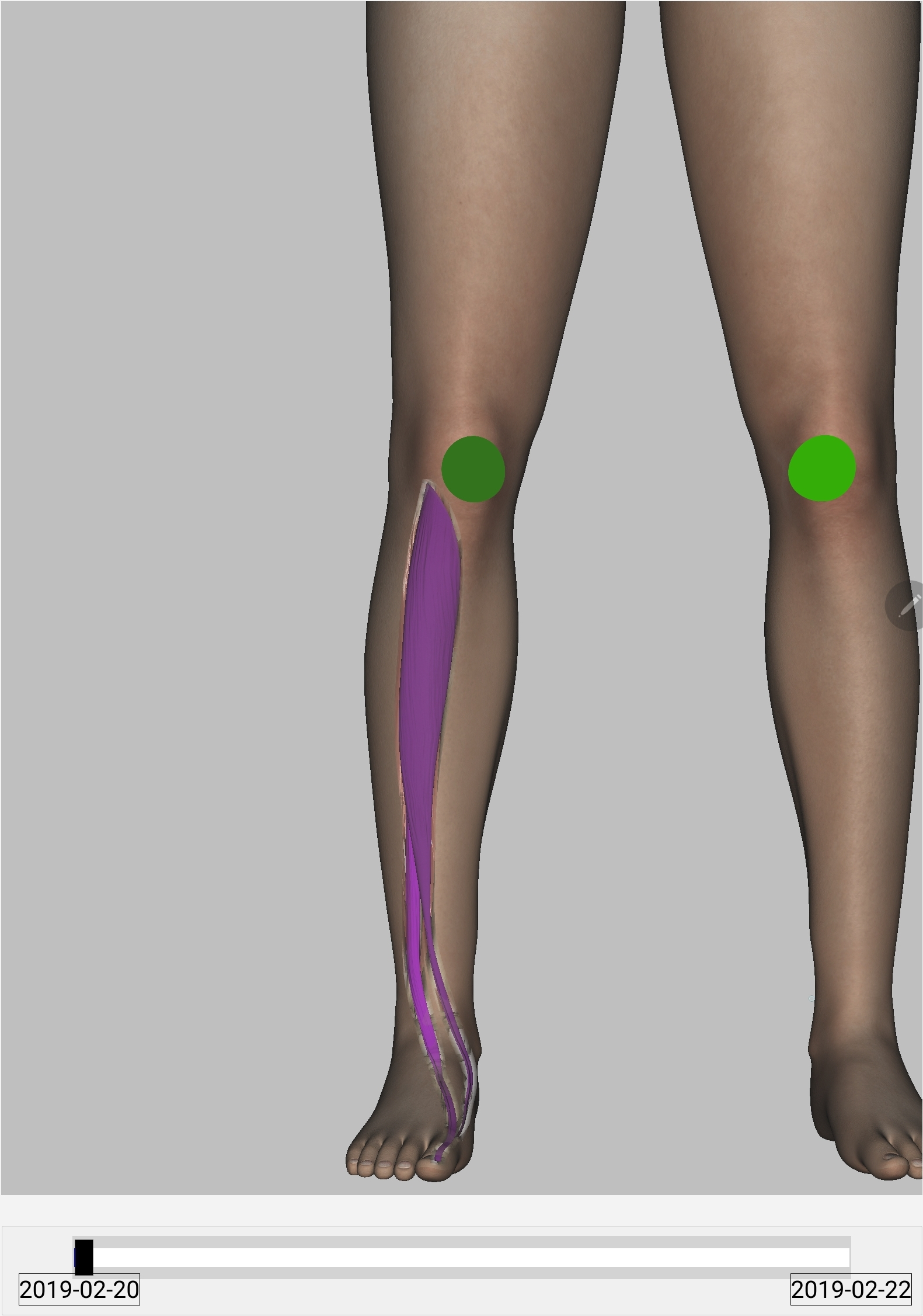} 
		\label{fig:task1Muscle}
	} 
	\subfloat[]{
		\includegraphics[width=0.32\textwidth]{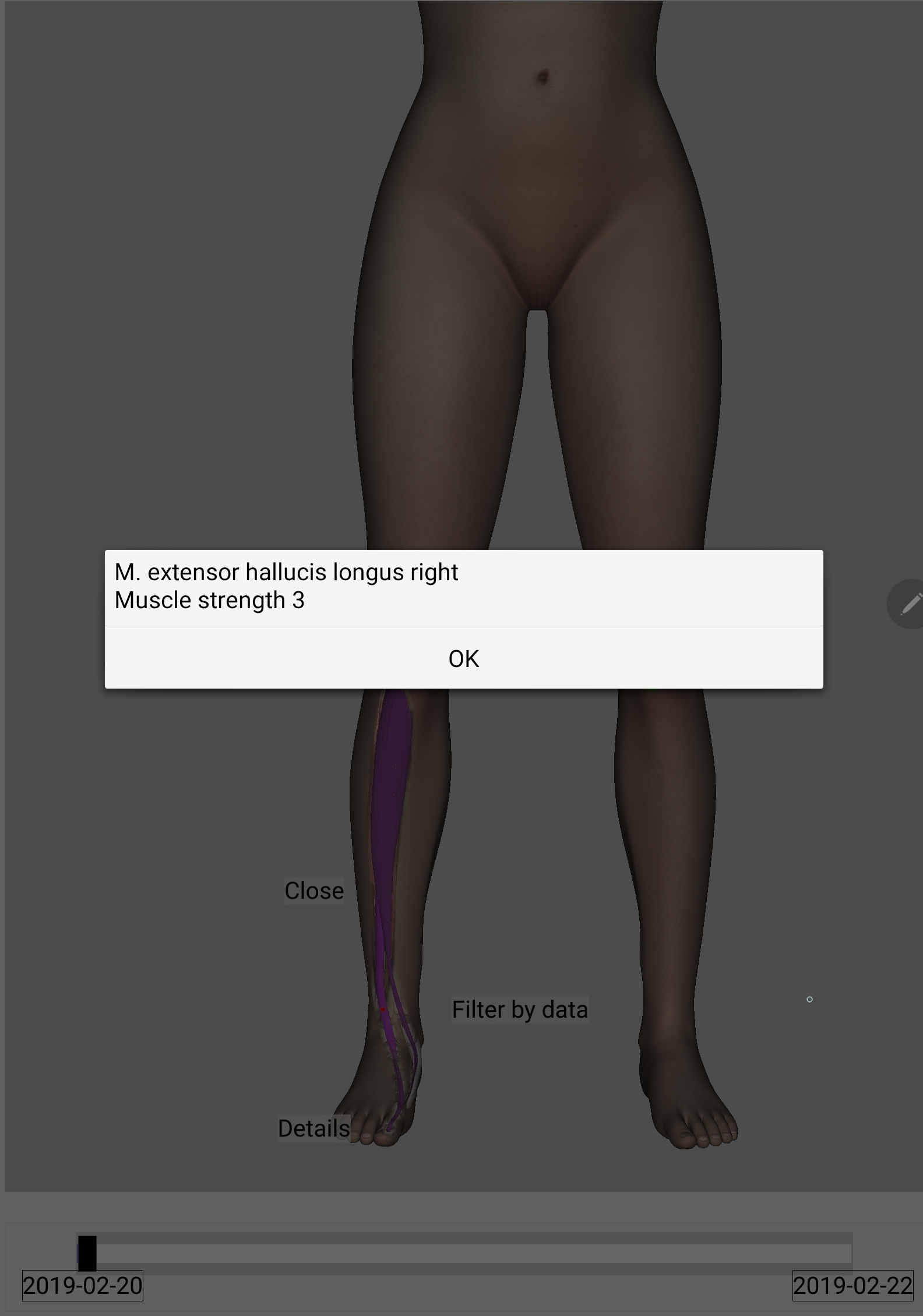} 
		\label{fig:task1Overlay}
	}
	\caption{Usage of further prototype features in \ref{enum:task2}. Fig.~\ref{fig:task1Filter} illustrates the use of the data category filter. Fig.~\ref{fig:task1Muscle} shows a close-up of paresis after filtering, \ie without the occluding dermatome. Fig.~\ref{fig:task1Overlay} exemplifies the textual overlay for one of the affected muscles comprising its anatomical name and associated raw muscle strength data.}
	\label{fig:task1Visibility}
\end{figure*}

\section{Prototype Evaluation}
\label{s:eval}

\subsection{Goals and Procedure}
\label{s:eval.proc}
The main evaluation objective is to verify the efficiency and effectiveness of the prototype concerning transfer of the cooperation relevant information in general, and its conformity with the design goals (see Sec~\ref{s:method.goals}) in particular. In order to test the aforementioned aspects, the evaluation is structured as follows.
\begin{enumerate*}[label=(\arabic*)]
	\item After a brief introduction to the overall context, the participants are directly confronted with \autoref{enum:task1}.
	\item Afterwards the legend is explained to the participants (see Fig.~\ref{fig:legend}) and they have to perform \autoref{enum:task2} and \autoref{enum:task3}.
	\item Finally, the participants have to reflect on the practical experience answering a questionnaire that comprises two main blocks, \ie \ref{enum:quest1} and \ref{enum:quest2}.
\end{enumerate*}

The three practical tasks are designed as follows:
\begin{enumerate}[itemindent=1.5cm, leftmargin=0cm,label=\textbf{Task~\arabic*.},ref=\textbf{Task~\arabic*}]
	\item \label{enum:task1} The goal of the first task is to assess the potential of our visualization approach for \textit{intuitive interpretability} (cf. \ref{enum:dg.intuitive}). Here the physicians need to give their spontaneous interpretation of two visualizations (see Fig.~\ref{fig:task0}) without previous knowledge about the meaning of the visual attributes.  
	
	\item \label{enum:task2} During the participatory refinement with the expert group (see Sec.~\ref{s:method}), we collected seven real-life descriptions of neurological statuses of patients with spinal disc herniation (cf.~\ref{enum:dg.synopsis}, \ref{enum:dg.concise}). All seven statuses are described in textual form similarly to the case presentations in the department morning meetings and one status is visualized with our prototype (see Fig.~\ref{fig:task1Alt1} and \ref{fig:task1Alt2}). In this task, the physicians have to \textit{assign the given visualization to the corresponding textual description}. A specific challenge arises from the fact, that the visualized status is very similar to another one in the list (cf. \ref{enum:dg.discriminability}). More precisely, the respective datasets differ only by the relatively small muscles in the right leg that are affected by paresis, which are anatomically adjacent, and both are partially occluded by the L5 dermatome.
	
	The given dataset allows to test the effectiveness of the prototype features such as \textit{alternating visualization} (representing the radicular pain and sensory disorder in same dermatome, in this case), \textit{dynamic occlusion handling} (over the affected muscles), \textit{proxy geometries} (for the patella tendons), and \textit{3D navigation}. Moreover, in order to solve this task correctly, the participants must also use the \emph{category filter}, which hides the occluding dermatome (see Fig.~\ref{fig:task1Filter} and \ref{fig:task1Muscle}), and, if necessary, activate the textual overlay (see Fig.~\ref{fig:task1Overlay}), \eg to identify the exact muscle strength level or muscle name.
	
	\item \label{enum:task3} The last task is to \textit{read the healing progress} (cf. \ref{enum:dg.progress}) from two visualizations that show successive patient states. The first one equals \ref{enum:task2} and  corresponds to the preoperative status of a patient. The second one represents the postoperative neurological status of the same patients. Switching between both visualizations by means of the timeline, the physicians have to recognize (cf. \ref{enum:dg.discriminability}) and to describe the respective changes. Analogously to \ref{enum:task2}, there is an additional level of difficulty due to the partially remaining paresis that is still occluded in the postoperative status (see Fig.~\ref{fig:task2Sensory}).
\end{enumerate}

The final questionnaire that is to be answered by the participants comprises the following two main blocks:
\begin{enumerate}[itemindent=13mm, leftmargin=0cm,label=\textbf{QNR~\arabic*.},ref=\textbf{QNR~\arabic*}]
	\item \label{enum:quest1} The first questionnaire is dedicated to the evaluation of the usage-dependent \textit{view} (cf. \ref{enum:dg.concise}). Particularly, the physicians assess the \textit{relevance of the pre-selected data categories} in the given cooperative settings as well as their \textit{completeness}.
	
	\item \label{enum:quest2} In the second block the physicians assess the benefit of the prototype for the transfer of information regarding a patient’s neurological status. Moreover, they are asked to explicitly specify advantages or drawbacks of the prototype's features, such as using the anatomy as spatial representation for data visualization (cf. \ref{enum:dg.familiar}), simultaneous visualization of multiple data (cf. \ref{enum:dg.synopsis}) and implementation on a mobile device (cf. \ref{enum:dg.mobility}).
\end{enumerate}

Ten neurosurgeons from the same department took part in the evaluation, none of which was involved in the participatory refinement process described in \autoref{s:method}, \ie this test group only had a very basic understanding of the overall aim of the visualization prototype. The group consists of three assistant physicians, one specialist, five senior physicians and one chief physician. The interviews were conducted in groups of two persons or, partially, individually, and have been filmed for documentation and analysis. The response rate of questionnaires has been nine out of ten.

\subsection{Results}
\label{s:eval.results}
In the following we present the main insights of the prototype evaluation, structured according to the aforementioned tasks.

The results of \ref{enum:task1} demonstrate that the anatomical integration (cf. \ref{enum:dg.familiar}) and the usage-dependent view (cf. \ref{enum:dg.concise} allow a rather intuitive interpratation by the physicians (cf.~\ref{enum:dg.intuitive}). They could successfully focus on the set of visual attributes related to the respective anatomical structure in the given context and interpret the underlying neurological information to a large extend w/o any pre-knowledge.
For instance, the purple color in the muscle (see Fig.~\ref{fig:task0_2}) was consistently associated with paresis (10/10) and the interpretation of the red dermatome (see Fig.~\ref{fig:task0_1}) varied between pain (8/10) and sensory disorder (2/10). At the same time, the data category \emph{tendon reflex} has often been misinterpreted (6/10). However, several participants rated this category as less relevant (see \ref{enum:quest1}), which is why it was less anticipated. In particular, the option `useful but dispensable' in the given view got by tendon reflex ca. 45\% of all votes, whereby the result of the respective questionnaire column for other categories vary between 11\% and 20\%. Moreover, the visualization of the \emph{tendon reflex} is anatomically weakly integrated due to the required proxies, which also affects its interpretability.
\newline	
To sum up, the prototype shows a considerable degree of intuitive comprehensibility (see \ref{enum:dg.intuitive}), which also was confirmed in the next tasks where the participants were able to successfully handle it after a very short learning phase. Therefore, we conclude that there is a strong relation between anatomical integration and view-dependent relevance of the data on the one hand, and intuitiveness of their visual interpretation on the other hand.

In \ref{enum:task2}, \ie the assignment of a visualization to textual descriptions, all physicians could easily (10/10) limit the conceivable variants to both cases: the correct case and one that was very similar, as described above. However, for the decision between both cases the most participants needed further hints, \eg to use the category filter and the zooming function. Notably, after receiving these hints, several participants (6/10) could familiarize themselves with the prototype features and independently apply them in  \ref{enum:task3} for the correct recognition of the remaining paresis.
\newline
In summary, the visualization prototype allows the physicians to successfully read the neurological status of a patient with spinal disk herniation (cf.~\ref{enum:dg.synopsis}, \ref{enum:dg.concise}). At the same time, a limited visibility of target anatomical structures, particularly several small targets structures close to each other (see Fig.~\ref{fig:task1Muscle}) or occlusions (see Fig.~\ref{fig:task1Alt1}-\ref{fig:task1Alt2}), can affect the discriminability of the respective visualization (cf. \ref{enum:dg.discriminability}) and consequently complicate its reading. Apparently, the users need a higher degree of familiarity to handle these problematic situations and to apply prototype features such as 3D navigation and category filtering. We will further investigate approaches with a more intuitive access to the synopsis of the patient's status (cf.~\ref{enum:dg.synopsis}, \ref{enum:dg.intuitive}) in these kinds of more complex situations.

The healing progress in \ref{enum:task3} (cf. \ref{enum:dg.progress}) was interpreted correctly by all participants (10/10). Most physicians also emphasized a high practical relevance of this kind of visualization. Note, that the correct interpretation of relative value changes, \ie improvement or deterioration, was possible without consulting the legend or calling the textual overlay. 

Moreover, several aspects addressed in the questionnaire blocks \ref{enum:quest1} and \ref{enum:quest1} have lead to additional insights discussed in the following.


The evaluation of \ref{enum:quest1} validated the \textit{relevance of the pre-selected data categories} for the specified view (cf. \ref{enum:dg.concise}), only the tendon reflex has been rated controversially. The latter is explained by individual variations in the clinical examination practices: while some physicians rarely apply the tendon reflex test, others consider it useful for obtaining additional insight into the patient's status. Regarding the \textit{completeness}, five physicians proposed to extend the view with further categories or properties, \eg 'pathological reflex' as category and 'duration' as type of a symptom property. These suggestions, again, mainly reflect differences in the individual professional procedures. Further suggestions, \eg 'previous surgeries', indicate the interest of using the prototype in further contexts by adding new views.

All, \ie nine out of nine, responses of \ref{enum:quest2} positively assess the potential of the visualization concept for enhancing and accelerating of the information transfer in the targeted cooperative situations. Especially, rating positively the simultaneous visualization (cf. \ref{enum:dg.synopsis}) and the use of the human anatomy (cf. \ref{enum:dg.familiar}), the physicians stated concrete advantages such as providing of \textit{multiple pieces of information at a glance} and consulting the model for \textit{refreshing their anatomical knowledge}. They also mentioned potential drawbacks in relation with these concepts such as a growing \textit{visualization complexity} by increase of the amount of data in a single view or a \textit{long learning curve}.

The physicians see the main contribution of the implementation on a mobile platform (cf. \ref{enum:dg.mobility}) in the possible \textit{time saving} thanks to the \textit{high availability} of patient data. At the same time, they expressed their concerns regarding the necessity to carry an additional device in their coat pocket.

In summary, it can be stated that the anatomically integrated in-place visualization of medical data was appreciated by the most participants: only one of them could not discern any benefit in the use of our approach in cooperative situations in comparison to textual data. In contrast, one neurosurgeon unsolicitedly drew the analogy between our prototype and the MR images that are used for cooperative surgery planning, as in both cases the specialists immediately ``see'' the relevant information. This clearly suggests, that the visualization concept can also be used in synchronous cooperative settings, for instance in the discussions during the department morning meetings.


\section{Conclusion and Future Work}
In this paper we presented a novel concept for anatomically integrated in-place visualization of medical data. The concept is designed in accordance with the requirements arising from specific tasks in cooperative clinical workflow, namely transfer of cooperation relevant neurosurgical information between colleagues. Our approach allows for a spatially integrated comprehensive visualization of abstract medical data, such as clinical symptoms, on a 3D human avatar using their inherent references to affected anatomical structures and an appropriate visual encoding. Preselecting patient data as a function of their relevance in the given clinical usage context, \ie view, provides an at-a-glance synopsis of relevant information to physicians.

The evaluation of the prototypical implementation of the visualization concept by a group of neurosurgeons revealed positive feedback, in particular concerning the use of anatomy as spatial representation of data and the potential speed-up of information assessment. At the same time, it revealed some limitations of the current solution in the situations where the target anatomical structures are not sufficiently distinguishable without user interaction.

Beyond the extension of our prototype to further clinical usage contexts, we intend to address several of the issues that have arisen in the evaluation, such as the flexibility to adapt to individual professional procedures and the problem in distinguishing adjacent anatomical structures.


\section*{Acknowledgments}
The authors would like to thank the medical staff at the neurosurgery department of the Jung-Stilling hospital for their willingness and engagement in our field and user studies. The work is funded by the German Research Foundation (DFG) in the context of the Collaboratice Research Center (SFB) 1187 ``Media of Cooperation'', sub-project A06 ``Visual integrated medical cooperation''.

\bibliographystyle{abbrv-doi}

\bibliography{references}
\end{document}